\RequirePackage{fix-cm}

\documentclass[smallextended]{svjour3}

\smartqed
\usepackage{graphicx}
\usepackage{subcaption} 
\usepackage{placeins}  
\usepackage{float}     
\usepackage{amsmath}

\usepackage[english]{babel}
\usepackage[numbers,sort&compress]{natbib}  
\usepackage[colorlinks=true, linkcolor=blue, citecolor=blue, urlcolor=blue]{hyperref}
\begin{document}

\title{Many-body localization properties of one-dimensional anisotropic spin-1/2 chains}
\subtitle{}

\titlerunning{Many-body localization properties of one-dimensional anisotropic spin-1/2 chains }
\author{Taotao Hu$^{\dag}$ \and Yuting Li   \and Jiameng Hong \and Xiaodan  Li  \and Dongyan Guo \and Kangning Chen}


\institute{Taotao Hu \and  Yuting Li
     \and
           Jiameng Hong
                       \and
            Dongyan Guo
               \and
           Kangnning Chen
         \at
              School of Physics, Northeast Normal University,
Changchun 130024, People's Republic of China\\
             \\
              \email{hutt262@nenu.edu.cn}           
           \and
           Xiaodan Li \at
             College of Science, University of Shanghai for Science and Technology, Shanghai 200093,  People's Republic of China}

\date{Received: date / Accepted: date}

\maketitle

\begin{abstract}

In this paper, we theoretically investigate the many-body localization (MBL) properties of one-dimensional anisotropic spin-1/2 chains by using the exact matrix diagonalization method. Starting from the Ising spin-1/2 chain, we introduce different forms of external fields and spin coupling interactions, and construct three distinct anisotropic spin-1/2 chain models. The influence of these interactions on the MBL phase transition is systematically explored.
We first analyze the eigenstate properties by computing the excited-state fidelity. The results show that MBL phase transitions occur in all three models, and that both the anisotropy parameter and the finite system size significantly affect the critical disorder strength of the transition. Moreover, we calculated the bipartite entanglement entropy of the system, and the critical points determined by the intersection of curves for different system sizes are basically consistent with those obtained from the excited-state fidelity. Then, the dynamical characteristics of the systems are studied through the time evolution of diagonal entropy (DE), local magnetization, and fidelity. These observations further confirm the occurrence of the MBL phase transition and allow for a clear distinction between the ergodic (thermal) phase and the many-body localized phase.
Finally, to examine the effect of additional interactions on the transition, we incorporate Dzyaloshinskii–Moriya (DM) interactions into the three models. The results demonstrate that the MBL phase transition still occurs in the presence of DM interactions. However, the anisotropy parameter and finite system size significantly affect the critical disorder strength. Moreover, the critical behavior is somewhat suppressed, indicating that DM interactions tend to inhibit the onset of localization.

\keywords{Many-body localization  \and One-dimensional spin-1/2 chain  \and Anisotropy  \and Dzyaloshinskii–Moriya  interactions  }
\end{abstract}
\clearpage

\section{Introduction}

In recent years, many-body localization (MBL) has emerged as a central topic in the study of non-equilibrium quantum systems. It reveals that in the presence of interactions and disorder, thermalization may fail, challenging the conventional eigenstate thermalization hypothesis (ETH)\cite{anderson1958absence,billy2008direct,
basko2006metal,safavi2019quantum,macieszczak2019coherence}. The MBL phase is characterized by the emergence of local integrals of motion (LIOMs), which inhibit entanglement spreading and enable the system to retain information about its initial conditions for arbitrarily long times, thus preventing it from reaching thermal equilibrium\cite{doggen2019many,
sierant2018many,
bardarson2012unbounded,
serbyn2013universal,
luitz2016extended,
pino2014entanglement,
deng2017logarithmic,
xu2018emulating,
torres2017extended}. Recent research has further highlighted the potential of MBL systems in preserving long-time quantum coherence even under periodic driving conditions, reinforcing their value in quantum information processing\cite{ponte2015periodically}. These findings strengthen the view that MBL provides a robust mechanism for long-term quantum information storage and memory\cite{bauer2013area,
huse2013localization,
rubin2019subnanometer}.

MBL was first extensively studied in one-dimensional spin-1/2 chains, with representative models including disordered Heisenberg and Ising chains\cite{pal2010many,
potter2015universal,
zhang2018universal,
canovi2011quantum,
serbyn2015criterion,
bairey2017driving,
schreiber2015observation,
bordia2016coupling}. As the disorder strength increases, the system undergoes a transition from an ergodic (thermalizing) phase to a non-thermal localized phase\cite{iyer2013many, khemani2017two}. In the ergodic regime, the system follows ETH and evolves toward thermal equilibrium, while in the localized regime, ETH breaks down and the eigenstates exhibit localization in Hilbert space\cite{lee2017many, devakul2015early}. 
These properties have made MBL a paradigmatic non-equilibrium quantum phase, extensively studied for its distinct dynamical behaviors. Various physical quantities, such as quantum Fisher information\cite{safavi2019quantum,macieszczak2019coherence,strek2015laser}, local magnetization\cite{ponte2015many,dutt2020higher}, and the power-law decay of imbalance\cite{chanda2020time,doggen2018many,doggen2019many,sierant2018many}, have been employed to explore the dynamical evolution of MBL systems. Among various indicators, the bipartite entanglement entropy (EE) plays a particularly important role in characterizing many-body phases, particularly in revealing the logarithmic spreading of entanglement in MBL regimes\cite{kjall2014many,luitz2015many,elliott2015many,bardarson2012unbounded,serbyn2013universal,luitz2016extended,pino2014entanglement,deng2017logarithmic}. However, the measurement of EE typically relies on quantum state tomography, which remains experimentally challenging in systems with many qubits\cite{xu2018emulating}.

To address these limitations, diagonal entropy (DE) has emerged as an efficient alternative. Studies have shown that DE also exhibits slow growth in MBL systems\cite{torres2017extended}, but it only requires the diagonal elements of the density matrix for its calculation, making it more accessible for both experimental and numerical investigations. Recent research further indicates that MBL can facilitate iterative quantum optimization, as demonstrated by the feasibility of constructing quantum approximate optimization algorithms (QAOA) based on the critical point of the MBL transition \cite{wang2022many,qu2018thermal}.Building on these theoretical and practical insights, it is now widely recognized that MBL not only serves as a fertile ground for exploring fundamental aspects of non-equilibrium quantum dynamics, but also holds considerable promise for advancing quantum information storage and algorithmic development.

Inspired by Anderson's theory of localization, researchers have attempted to introduce disorder into quantum systems with many-body interactions to induce MBL. It has been shown that uniform random disorder can trigger MBL transitions in a variety of models\cite{pal2010many,wang2022many}. In particular, studies on the disordered Ising model demonstrate that strong disorder can localize the system's eigenstates, thereby suppressing thermalization and enabling the system to preserve initial-state information for long times\cite{geng2020many}.Despite extensive research on MBL, most work has focused on disordered Ising and Heisenberg models in one dimension, and there has been relatively less research on many-body localization in one-dimensional anisotropic spin-1/2 chains. Therefore, in this study, we systematically investigate the effects of different interaction terms and external fields on the MBL phase transition in this model using exact diagonalization. We explore how the system undergoes a transition from thermalization to localization under different parameter conditions. The study shows that external fields and interaction strengths play a crucial role in the MBL phase transition, influencing the system's localization properties and providing a more detailed understanding of the many-body localization phase transition.

\section{NUMERICAL MODEL}


On the basis of one-dimensional Ising spin-1/2 chains we study three models: spin-1/2 chains with each anisotropic next-nearest-neighbor coupling, spin-1/2 chains with each anisotropic external field, and spin-1/2 chains with each anisotropic all non-nearest-neighbor coupling.

The Hamiltonian of the one-dimensional Ising spin-1/2 chain with next-nearest-neighbor couplings in the spin-$x$ and spin-$y$ directions can be expressed as follows:
\begin{equation}\label{eq2.7}
\hat{H}_1 = \sum\limits_{i = 1}^{N - 1}  \hat{S}_i^z  \hat{S}_{i + 1}^z + \sum\limits_{i = 1}^N {h_i} \hat{S}_i^z + \sum\limits_{i = 1}^{N - 2} [(1 + \gamma)  \hat{S}_i^x  \hat{S}_{i + 2}^x + (1 - \gamma)  \hat{S}_i^y  \hat{S}_{i + 2}^y ],
\end{equation}

The Hamiltonian of the one-dimensional Ising spin-1/2 chain with a constant external field applied in the spin-$x$ direction can be expressed as follows:
\begin{equation}
\hat{H}_2 = \sum\limits_{i = 1}^{N - 1}  \hat{S}_i^z  \hat{S}_{i + 1}^z + \sum\limits_{i = 1}^N {h_i} \hat{S}_i^z +\gamma  \sum\limits_{i = 1}^{N }  \hat{S}_i^x,
\end{equation}

The Hamiltonian of the one-dimensional Ising spin-1/2 chain with all non-nearest-neighbor couplings in the spin-$x$ and spin-$y$ directions can be expressed as follows:
\begin{equation}
\hat{H}_3 = \sum\limits_{i = 1}^{N - 1}  \hat{S}_i^z  \hat{S}_{i + 1}^z + \sum\limits_{i = 1}^N {h_i} \hat{S}_i^z +\sum_{i=1}^{N-2} \sum_{j>i+1}^{N} \frac{1}{(j-i)^{\gamma}}\left[(1-\gamma) \hat{S}_{i}^{y} \hat{S}_{j}^{y}+(1+\gamma)  \hat{S}_{i}^{x} \hat{S}_{j}^{x}\right],
\end{equation}

where the  $S_i^x$, $S_i^y$, and $S_i^z$ are the spin-1/2 operators, $h_i$ is the disordered external field at site 
$i$, randomly distributed within the range $[-h,h]$. 
$\gamma$ is the anisotropy parameter, in the third model, $\gamma$ also affects the spin interaction range of the system.

\section{Results and Discussion}
Fidelity is a key tool in quantum information theory for measuring the similarity between quantum states and is widely used in the study of quantum phase transitions. In many-body localized (MBL) phase transitions, the fidelity can reveal the transition of the system from an ergodic phase to a localized phase.By calculating the fidelity, it is possible to effectively capture the characteristics of the MBL phase transition\cite{zanardi2007information,
albuquerque2010quantum,
hu2016fidelity,
hu2017excited,
sun2019fidelity}. In this paper, we use excited state fidelity to study the MBL phase transition. According to reference \cite{zanardi2006ground}, the fidelity of the $n$th excited state $\psi_n(h)$ in the system is defined as the overlap between excited states with parameters $h$ and $h+\delta h$:

\begin{equation}
F_n(h, h + \delta h) = \left| \langle \psi_n(h) | \psi_n(h + \delta h) \rangle \right|,
\end{equation}

For the tiny perturbation $\delta h_i$ at each site, we set $\delta h_i=\epsilon h_i$, where $\epsilon h_i=10^{-3}$. Special attention should be paid to the fact that the tiny perturbations $\delta h_i$ on different lattice points are not the same, but are independent random variables. In general, the fidelity value is close to 1, but near the critical point of phase transition, there will be significant changes in the fidelity value.

To better characterize the MBL phase transition, we selected excited states located in the middle third of the energy spectrum, since the fidelity of high-energy states provides a more pronounced indication of localization behavior\cite{vznidarivc2008many}.  We then compute the fidelity $F_n$ for each eigenstate $|\psi_n\rangle$. Averaging over all selected excited states and disorder realizations yields the mean value $\mathsf{E}[F]$. For each disorder amplitude $h$, we used 5000 disorder realizations for $N = 6$, 3000 realizations for $N = 8$ and 1000 realizations for $N = 10$ in this paper. Then we use exact matrix diagonalization for numerical analysis to obtain the data in this paper. The parallel programming techniques were employed to make computations feasible.

\begin{figure}
    \centering
    \begin{subfigure}[b]{0.50\textwidth}
      \includegraphics[width=\textwidth]{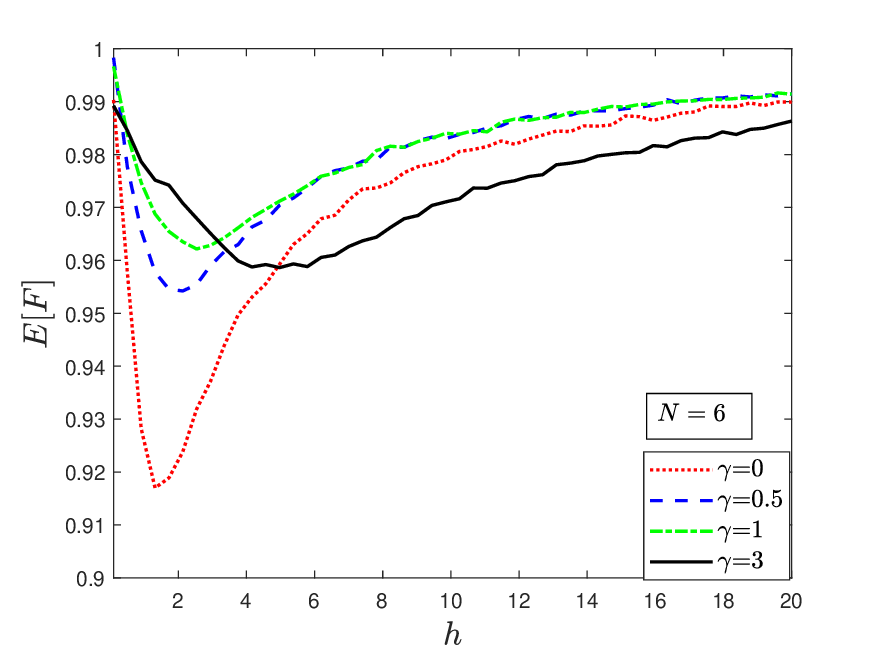}
      \caption{}
      \label{fig:oaspl_a}
    \end{subfigure}%
    ~
    \begin{subfigure}[b]{0.50\textwidth}
      \includegraphics[width=\textwidth]{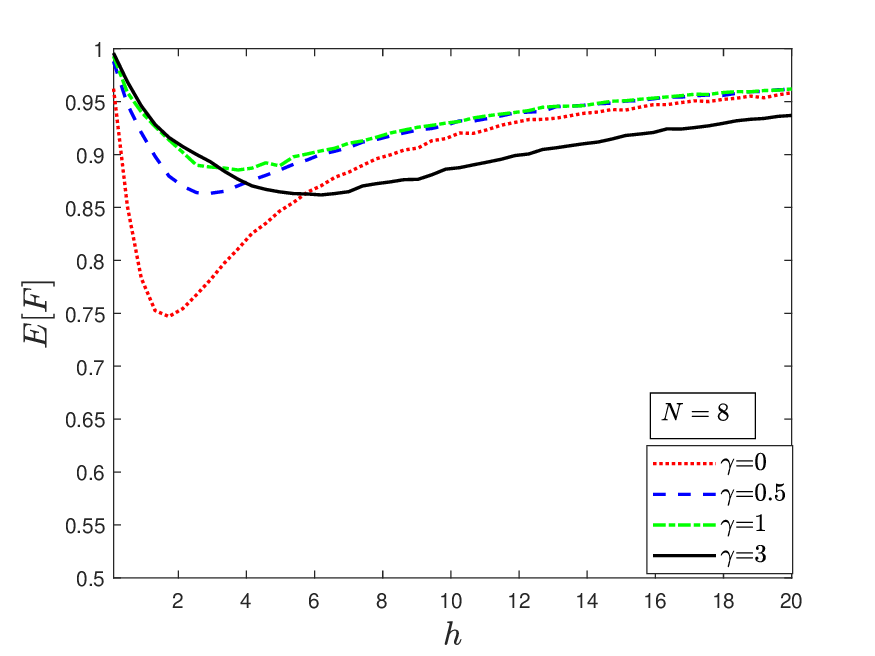}
      \caption{}
      \label{fig:oaspl_b}
    \end{subfigure} 
    
    \caption{The averaged excited-state fidelity $\mathsf{E}[F]$ as a function of disorder strength $h$ for the anisotropic spin-1/2 chain with anisotropic next-nearest-neighbor coupling at different values of the anisotropy parameter $\gamma$. (a) System size $N=6$; (b) System size $N=8$. The legend indicates the specific values of $\gamma$.}
   \label{fig1}
    \end{figure}
    
\begin{figure}
    \centering
    \begin{subfigure}[b]{0.50\textwidth}
      \includegraphics[width=\textwidth]{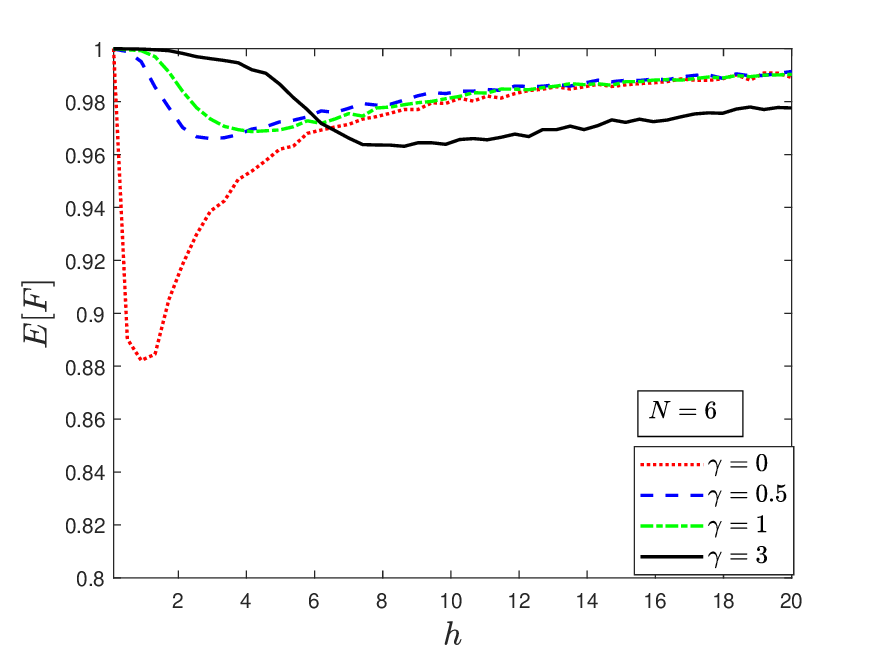 }
      \caption{}
      \label{fig:oaspl_a}
    \end{subfigure}%
    ~
    \begin{subfigure}[b]{0.50\textwidth}
      \includegraphics[width=\textwidth]{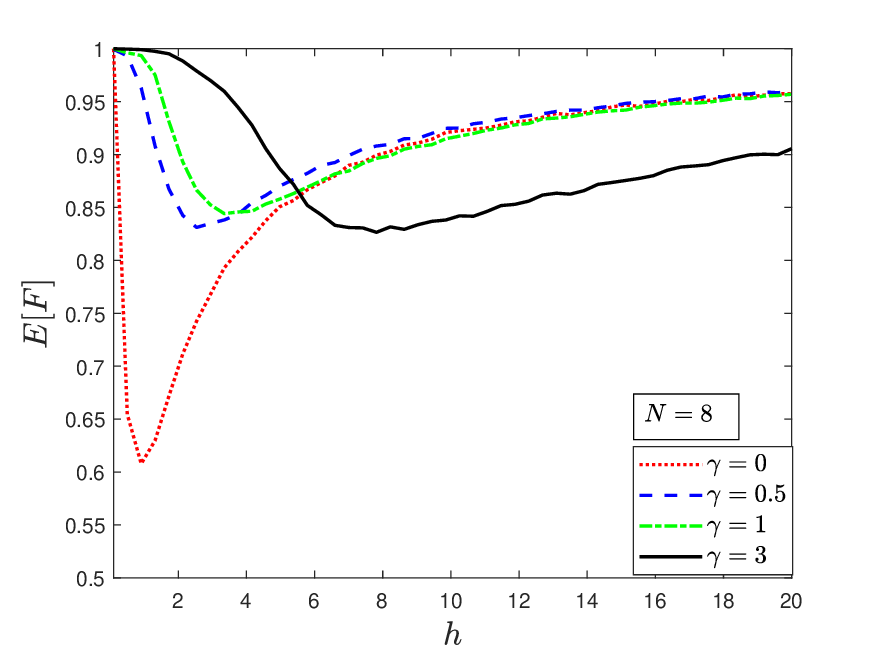}
      \caption{}
      \label{fig:oaspl_b}
    \end{subfigure} 
    
    \caption{The averaged excited-state fidelity $\mathsf{E}[F]$ as a function of disorder strength $h$ for the anisotropic spin-1/2 chain with anisotropic external field at different values of the anisotropy parameter $\gamma$. (a) System size $N=6$; (b) System size $N=8$. The legend indicates the specific values of $\gamma$.}
   \label{fig2}
    \end{figure}

 \begin{figure}
    \centering
    \begin{subfigure}[b]{0.50\textwidth}
      \includegraphics[width=\textwidth]{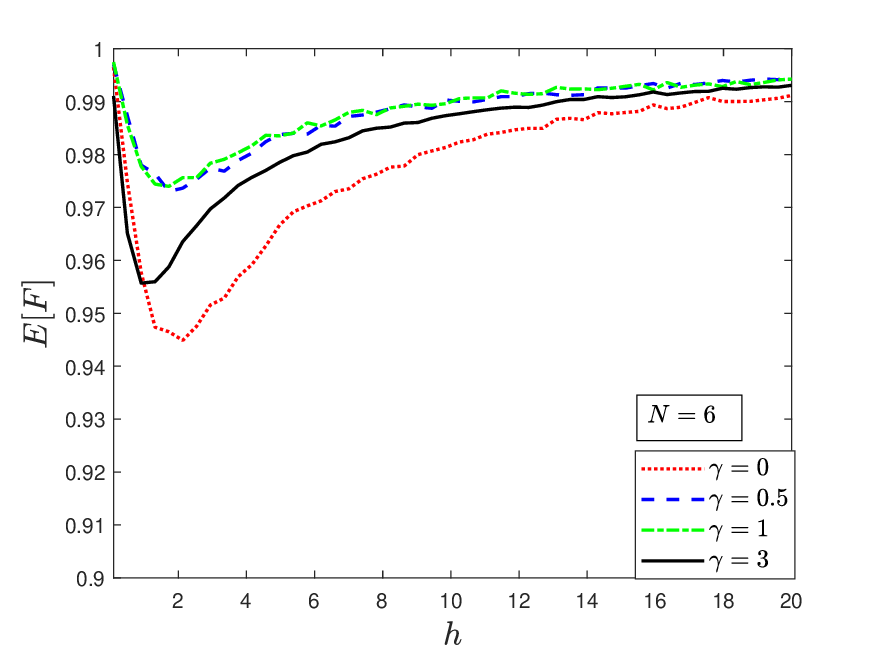 }
      \caption{}
      \label{fig:oaspl_a}
    \end{subfigure}%
    ~
    \begin{subfigure}[b]{0.50\textwidth}
      \includegraphics[width=\textwidth]{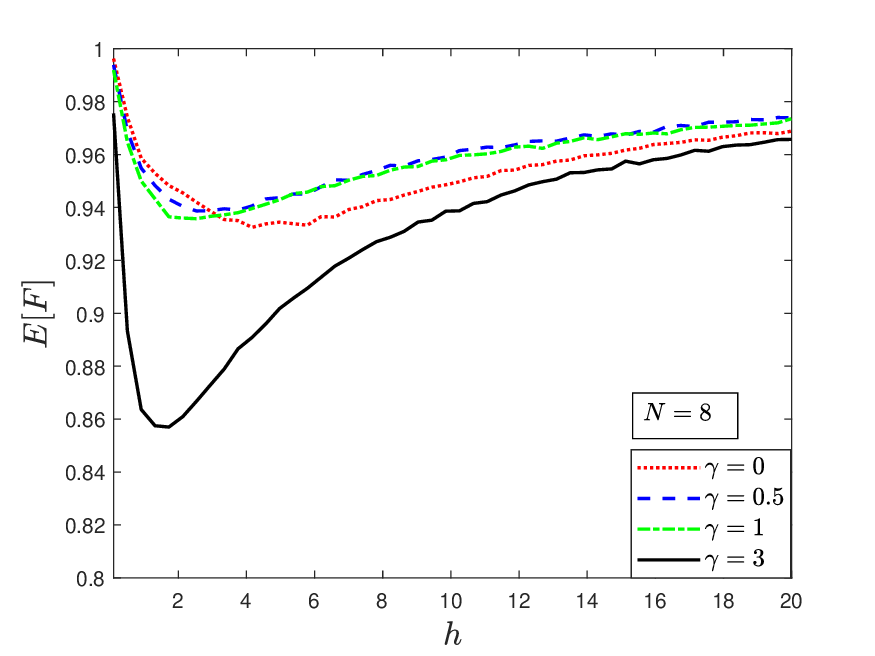}
      \caption{}
      \label{fig:oaspl_b}
    \end{subfigure} 
    
    \caption{The averaged excited-state fidelity $\mathsf{E}[F]$ as a function of disorder strength $h$ for the anisotropic spin-1/2 chain with anisotropic all non-nearest-neighbor couplings at different values of the anisotropy parameter $\gamma$. (a) System size $N=6$; (b) System size $N=8$. The legend indicates the specific values of $\gamma$.}
   \label{fig3}
    \end{figure}
    
We first investigated the effect of the anisotropy parameter $\gamma$ on the MBL phase transition in three one-dimensional anisotropic spin-1/2 chains, with system sizes $N = 6$ and $N = 8$. To characterize the transition behavior, we plotted the averaged excited-state fidelity $\mathsf{E}[F]$ as a function of disorder strength $h$, as shown in Fig. 1, Fig. 2, and Fig. 3. Based on previous studies\cite{wang2022many,qu2018thermal}, we can roughly get the critical point of the MBL phase transition from the inflection points of fidelity curves in excited states of the MBL phase transition. As seen in the three figures, increasing the disorder strength $h$ causes the excited-state fidelity $\mathsf{E}[F]$ to gradually decrease from an initial value close to 1. When $h$ approaches the critical value $h_c$, $\mathsf{E}[F]$ reaches a minimum. As $h$ continues to increase beyond $h_c$, the fidelity begins to rise again and eventually stabilizes at a constant value.

The results show that all three anisotropic spin-1/2 models undergo the MBL phase transition. However, the critical points of the MBL transition are affected by the anisotropy parameters. In addition, the impact of anisotropy differs among the three models, indicating that the response of each system to the variation in anisotropy is not completely the same. In the anisotropic spin-1/2 chains with anisotropic next-nearest-neighbor coupling and anisotropic external field, the critical point of the MBL phase transition increases with the anisotropy parameter $\gamma$. However, in the spin-1/2 chain with anisotropic all non-nearest-neighbor couplings, the critical point decreases as $\gamma$ increases, and the rate of decrease gradually slows down.
\begin{figure}
\centering
\includegraphics[width=0.70\textwidth]{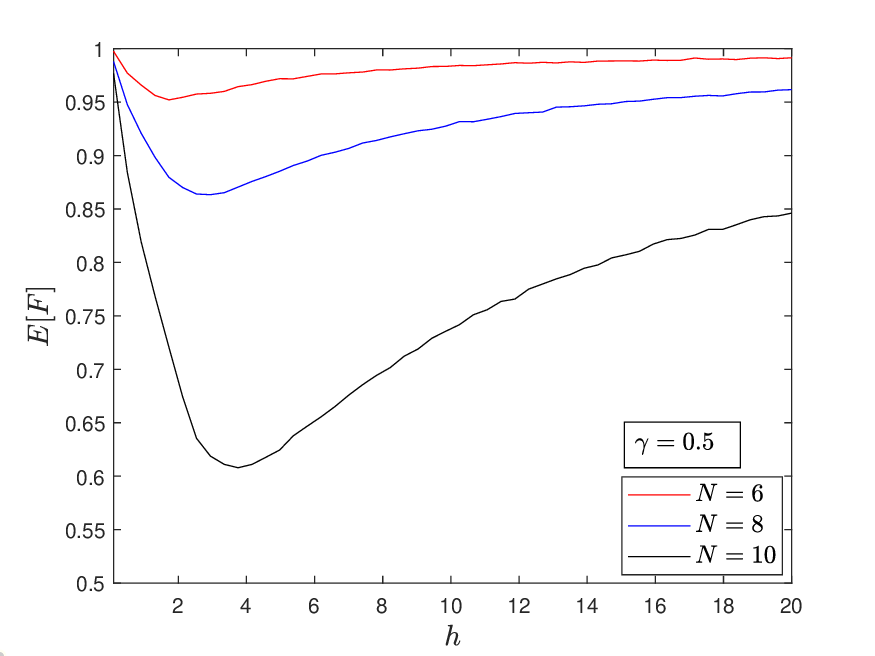}
\caption{ The averaged excited-state fidelity $\mathsf{E}[F]$ as a function of disorder strength $h$ for the anisotropic spin-1/2 chain with anisotropic next-nearest-neighbor coupling at a fixed anisotropy parameter $\gamma = 0.5$. The legend indicates the system sizes.}
\label{fig4}
\end{figure}

\begin{figure}
\centering
\includegraphics[width=0.70\textwidth]{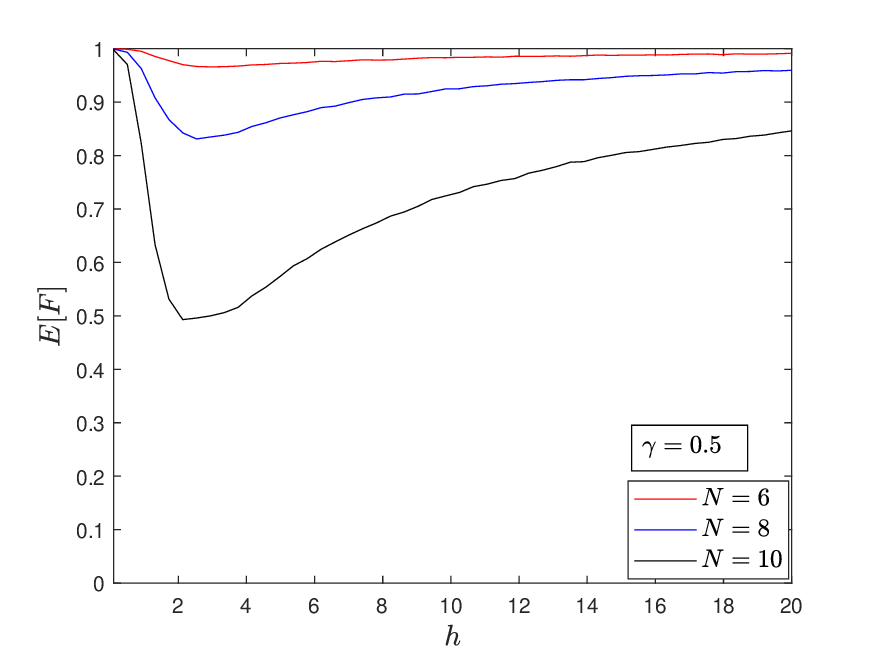}
\caption{ The averaged excited-state fidelity $\mathsf{E}[F]$ as a function of disorder strength $h$ for the anisotropic spin-1/2 chain with an anisotropic external field at a fixed anisotropy parameter $\gamma = 0.5$. The legend indicates the system sizes.}
\label{fig5}
\end{figure}

\begin{figure}
\centering
\includegraphics[width=0.70\textwidth]{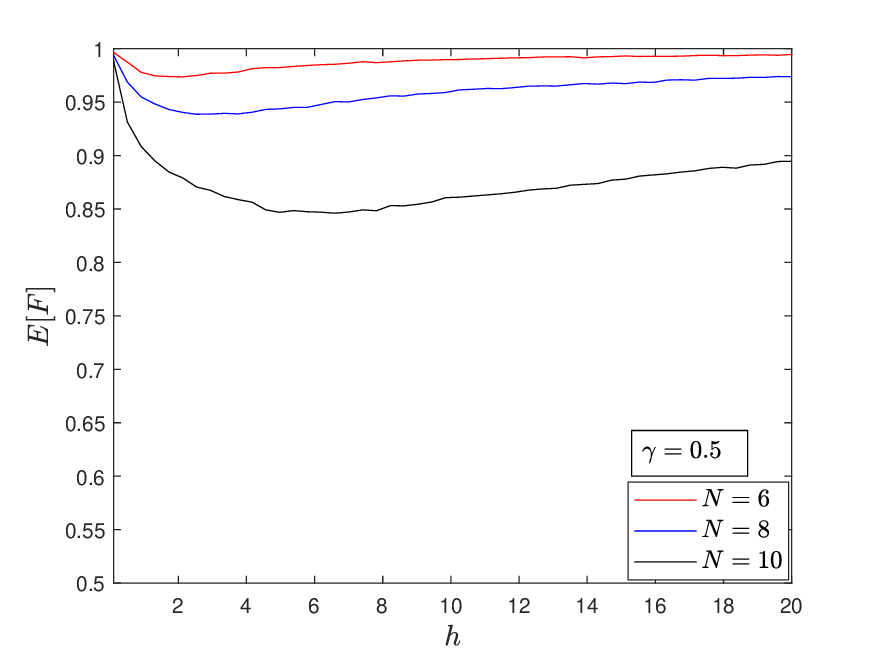}
\caption{ The averaged excited-state fidelity $\mathsf{E}[F]$ as a function of disorder strength $h$ for the anisotropic spin-1/2 chain with anisotropic all non-nearest-neighbor couplings at a fixed anisotropy parameter $\gamma = 0.5$. The legend indicates the system sizes.}
\label{fig6}
\end{figure}

Then we investigated the localization properties of many-body systems with finite system sizes, focusing on the effect of system size on the MBL phase transition of the three models at a fixed anisotropy parameter $\gamma = 0.5$. As shown in Fig. 4, Fig. 5, Fig. 6, we plotted the averaged excited-state fidelity $\mathsf{E}[F]$ as a function of disorder strength $h$ for system sizes $N = 6$, $8$,  $10$.

The three figures indicate that the critical point of the MBL phase transition in one-dimensional anisotropic spin-1/2 chains is dependent on the finite system size. For spin-1/2 chains with anisotropic next-nearest-neighbor coupling and with anisotropic all non-nearest-neighbor couplings, the critical point increases $h_c$ as the system size $N$ becomes larger, as seen in Fig. 4 and Fig. 6. In contrast, for the spin-1/2 chain with an anisotropic external field, the critical point $h_c$ slightly decreases with increasing system size $N$, as shown in Fig. 5.

\begin{figure}
    \centering
     \includegraphics[width=0.70\textwidth]{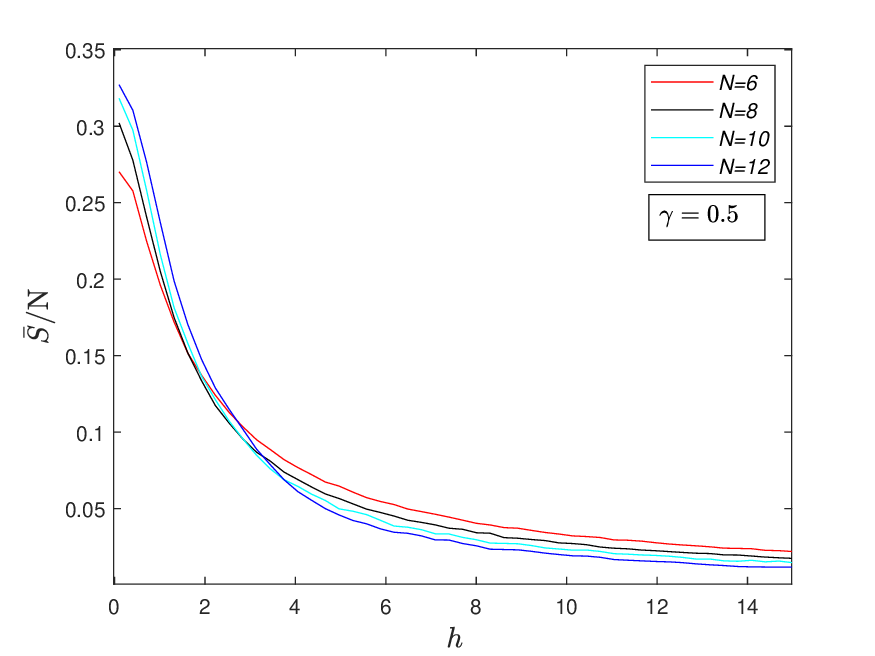}
   \caption{The averaged bipartite entanglement entropy over the system size $\bar{S}/N$ for the anisotropic spin-1/2 chain with anisotropic next-nearest-neighbor couplings as a function of the disorder strength $h$ at a fixed anisotropy parameter $\gamma = 0.5$. The system sizes are indicated in the legend.}
    \label{fig7}
    \end{figure}
\begin{figure}
    \centering
     \includegraphics[width=0.70\textwidth]{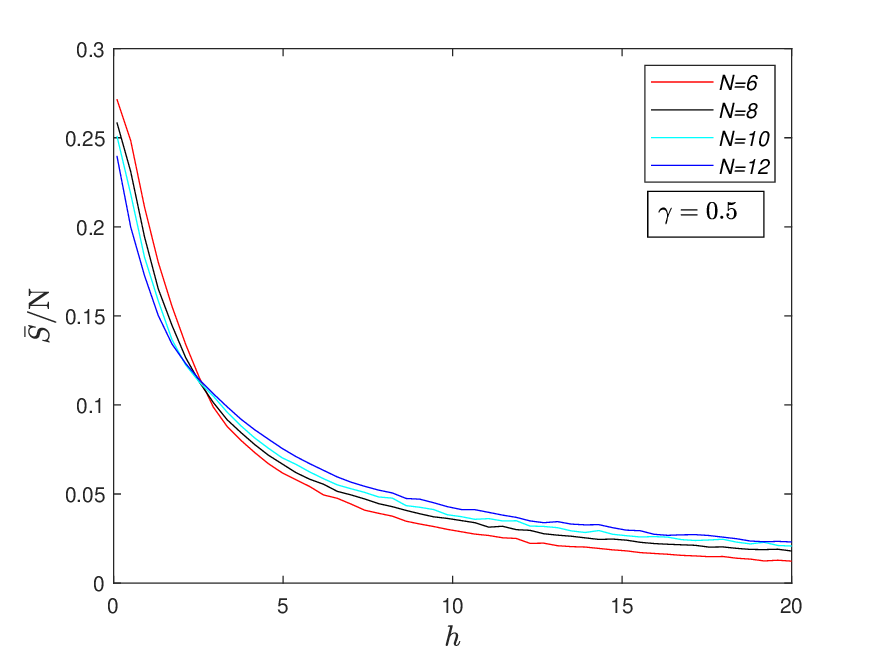}
   \caption{The averaged bipartite entanglement entropy over the system size $\bar{S}/N$ for the anisotropic spin-1/2 chain with anisotropic external fields as a function of the disorder strength $h$ at a fixed anisotropy parameter $\gamma = 0.5$. The system sizes are indicated in the legend.}
    \label{fig8}
    \end{figure}    
\begin{figure}
    \centering
     \includegraphics[width=0.70\textwidth]{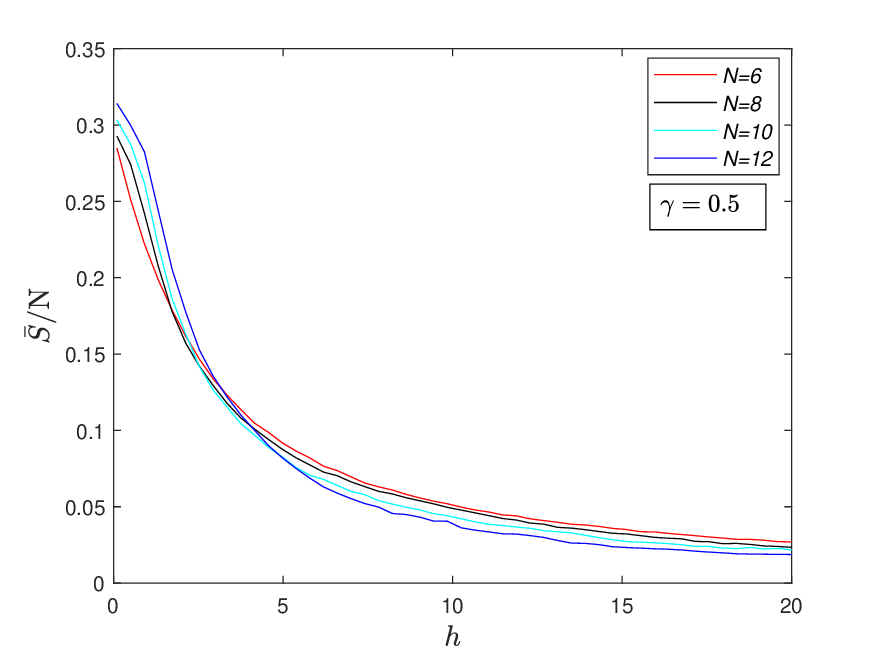}
   \caption{The averaged bipartite entanglement entropy over the system size $\bar{S}/N$ for the anisotropic spin-1/2 chain with anisotropic all non-nearest-neighbor couplings as a function of the disorder strength $h$ at a fixed anisotropy parameter $\gamma = 0.5$. The system sizes are indicated in the legend.
    }
    \label{fig9}
    \end{figure}

To demonstrate that the inflection point of the excited-state fidelity curve can indicate the critical point of the MBL phase transition, we calculate the bipartite entanglement entropy for the three models. We divide the one-dimensional chain into two equal parts, denoted as subsystems $A$ and $B$, each of length $N/2$. Then we compute the von Neumann entanglement entropy of subsystem $A$, based on its reduced density matrix $\rho_A$, defined as\cite{kjall2014many,luitz2015many,elliott2015many}:

\begin{equation}
S = -\mathrm{Tr}(\rho_A \ln \rho_A),
\end{equation}

where $\rho_A = \mathrm{Tr}_B(|\psi_n\rangle \langle \psi_n|)$ is the reduced density matrix of the left half of the system, and $|\psi_n\rangle$ is the $n$-th excited eigenstate of the full system. For system sizes $N = 6, 8, 10, 12$, we perform 1000, 800, 500, and 300 realizations of disorder, respectively. We compute the entanglement entropy for all selected excited eigenstates across all disorder realizations and take the averaged to obtain the mean bipartite entanglement entropy $\bar{S}$. By varying the disorder strength $h$, we can detect possible phase transitions and extract the critical point from the behavior of the entanglement entropy\cite{kjall2014many,luitz2015many,elliott2015many}.

Figures 7, 8, and 9 respectively illustrate the behavior of the normalized bipartite entanglement entropy $\bar{S}/N$ as a function of the disorder strength $h$ for three different models: the anisotropic next-nearest-neighbor coupled model, the model with anisotropic external fields, and the model with all anisotropic non-nearest-neighbor couplings. All simulations are performed with a fixed anisotropy parameter $\gamma = 0.5$. In the weak disorder regime, increasing the system size $N$ leads to an increase in $\bar{S}/N$, consistent with the volume law and indicating strong quantum entanglement. As $h$ increases, the curves exhibit pairwise intersections, and eventually, $\bar{S}/N$ sharply drops close to zero, suggesting that the system enters a localized phase with significantly reduced entanglement in its eigenstates.

For these three models, the critical point of the MBL phase transition can be analyzed through the intersection points of the curves for adjacent system sizes. The results show that the critical point exhibits different dependencies on the system size. Specifically:

\begin{itemize}
  \item For the model with anisotropic next-nearest-neighbor couplings, the critical point is approximately $h_c \approx 1.6$ for system size $N = 6$ (the intersection point of the curves for system sizes $N = 6$ and $8$), $h_c \approx 2.8$ for $N = 8$  (the intersection point of the curves for system sizes $N = 8$ and $10$), and $h_c \approx 3.7$ for $N = 10$  (the intersection point of the curves for system sizes $N = 10$ and $12$), showing an increasing trend with system size.
  \item For the model with anisotropic external fields, the intersections occur around $h_c \approx 2.7$, $2.5$, and $2.0$, suggesting a slight decrease in the critical point as the system size increases.
  \item For the model with all anisotropic non-nearest-neighbor couplings, the corresponding critical points are approximately $h_c \approx 1.7$, $2.5$, and $4.8$, also exhibiting an increasing trend.
\end{itemize}

These critical values are in good agreement with those obtained from the excited-state fidelity analysis, confirming the effectiveness of bipartite entanglement entropy as a diagnostic tool for identifying MBL phase transitions. In conclusion, all three one-dimensional anisotropic spin-1/2 models exhibit a transition from a thermalized phase to a many-body localized phase, and the critical points can be approximately determined through finite-size scaling of entanglement entropy.

To further distinguish between the ergodic phase and the many-body localized phase, we next investigate the dynamical properties of the three models. Assuming the initial state of the system is the N\'eel state $|\psi(0)\rangle$, after time evolution under the unitary operator $U = e^{-iHt}$, the quantum state at time $t$ becomes $|\psi(t)\rangle = e^{-iHt} |\psi(0)\rangle$. The corresponding density matrix at time $t$ is given by $\rho(t) = |\psi(t)\rangle\langle\psi(t)|$.
We first calculate the dynamics of diagonal entropy (DE) with disordered external field, the DE has the form\cite{torres2017extended}:

\begin{equation}
S^{{\rm diag}}(t)=-{\rm Tr}(\rho^{\text{diag}}(t){\rm ln}\rho^{\text{diag}}(t)),
\end{equation}

where $\rho_{\mathrm{diag}}(t)$ denotes the diagonal part of the density matrix $\rho(t)$ at time $t$, which represents the probability distribution of the system in the eigenbasis.Averaging over all disorder realizations yields the mean value $\mathsf{E}[S^{{\rm diag}}(t)]$ 5000 disorder realizations for $N=6$ is used here.

From Fig. 4, we determine that for the anisotropic spin-1/2 chain with next-nearest-neighbor coupling at $\gamma = 0.5$ and $N = 6$, the critical disorder strength is approximately $h_c \approx 1.7$. Thus, Fig. 10 presents the long-time dynamics of DE for different disorder strengths under the same conditions.

\begin{figure}
\centering
\includegraphics[width=0.70\textwidth]{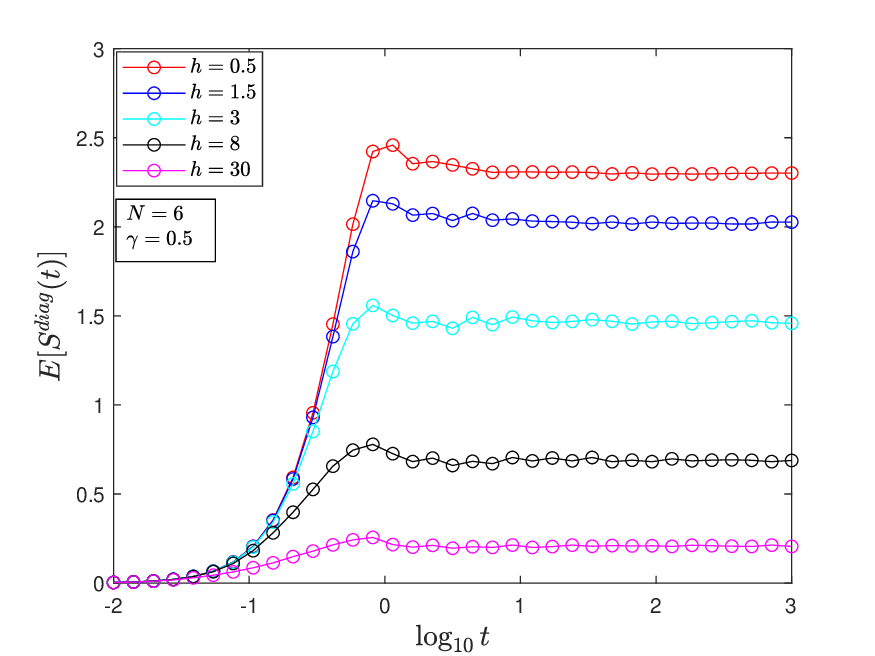}
\caption{ The dynamical growth of the averaged diagonal entropy $\mathsf{E}[S^{\mathrm{diag}}(t)]$ for the anisotropic spin-1/2 chain with anisotropic next-nearest-neighbor coupling at system size $N=6$ and anisotropy parameter $\gamma=0.5$. The legend indicates different values of disorder strength $h$.}
\label{fig10}
\end{figure}

As depicted in Fig. 10, at low disorder strength ($h = 0.5$), the DE rapidly increases, indicating fast quantum information spreading and quick thermalization in the ergodic phase. Over time, DE saturates at a high value, signifying thermal equilibrium.
At high disorder strength ($h = 30$), DE grows much more slowly, reflecting restricted information spreading and suppressed thermalization in the localized phase. Eventually, it saturates at a lower value, indicating local equilibrium with inhibited global thermalization.

We continue our investigation into the evolution of local magnetization, computing the expectation value of
the spin on a given site $I$ at time $t$ to obtain the magnetization $M_I(t)$, expressed as\cite{pal2010many}:
\begin{equation}
 M_I(t)=\langle\psi(t)\mid\hat{S_I^z}\mid\psi(t)\rangle,
\end{equation}

Averaging over all disorder realizations yields the mean
value $\mathsf{E}[M_I(t)]$, 5000 disorder realizations for $N = 6$ is used here. Fig. 11 illustrates the dynamical decay of the averaged local magnetization $E[\mathrm{MI}(t)]$ for the anisotropic spin-1/2 chain with an anisotropic external field at system size $N = 6$ and anisotropy parameter $\gamma = 0.5$, under different disorder strengths $h$.

\begin{figure}
\centering
\includegraphics[width=0.70\textwidth]{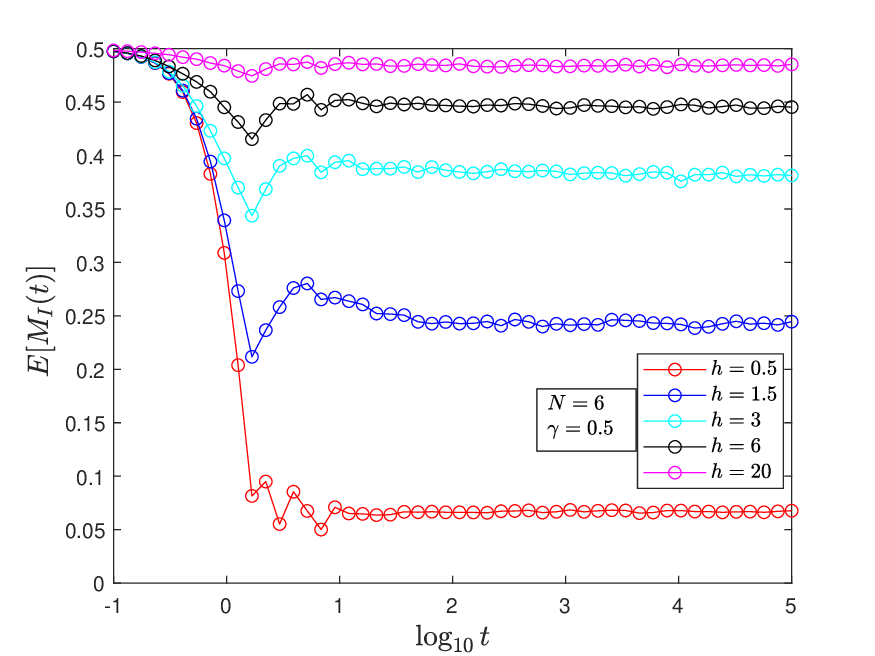}
\caption{The dynamical decay of the averaged local magnetization $\mathsf{E}[M_I(t)]$ for the anisotropic spin-1/2 chain with an anisotropic external field at system size $N = 6$ and anisotropy parameter $\gamma = 0.5$. The legend indicates different values of disorder strength $h$.}
\label{fig11}
\end{figure}

At low disorder strength ($h = 0.2$), the system rapidly thermalizes, with magnetization gradually decaying and stabilizing at a low value, indicating the ergodic phase. At high disorder strength ($h = 30$), magnetization remains around $1/2$, signifying the localized phase. In this regime, long-range order is destroyed, information propagation is restricted, and quantum states become confined to local regions.

Next we also study the time evolution of Fidelity $F(t)$, for mixed states, it is defined as\cite{wang2022many},
\begin{equation}
 F(t)={\rm Tr}[\rho(t)^{1/2}\rho_0\rho(t)^{1/2}]^{1/2},
\end{equation}
where $\rho(t) = |\psi(t)\rangle \langle \psi(t)|$, $\rho(0) = |\psi(0)\rangle \langle \psi(0)|$. 
Averaging over all disorder realizations yields the mean value $\mathsf{E}[F(t)]$. 
Here, 5000 disorder realizations are used for $N = 6$.
Fig. 12 illustrates the dynamical evolution of the averaged excited-state fidelity $\mathsf{E}[F(t)]$ for the anisotropic spin-1/2 chain with anisotropic all non-nearest-neighbor couplings at system size $N = 6$ and anisotropy parameter $\gamma = 0.5$, under different disorder strengths $h$.

\begin{figure}
\centering
\includegraphics[width=0.70\textwidth]{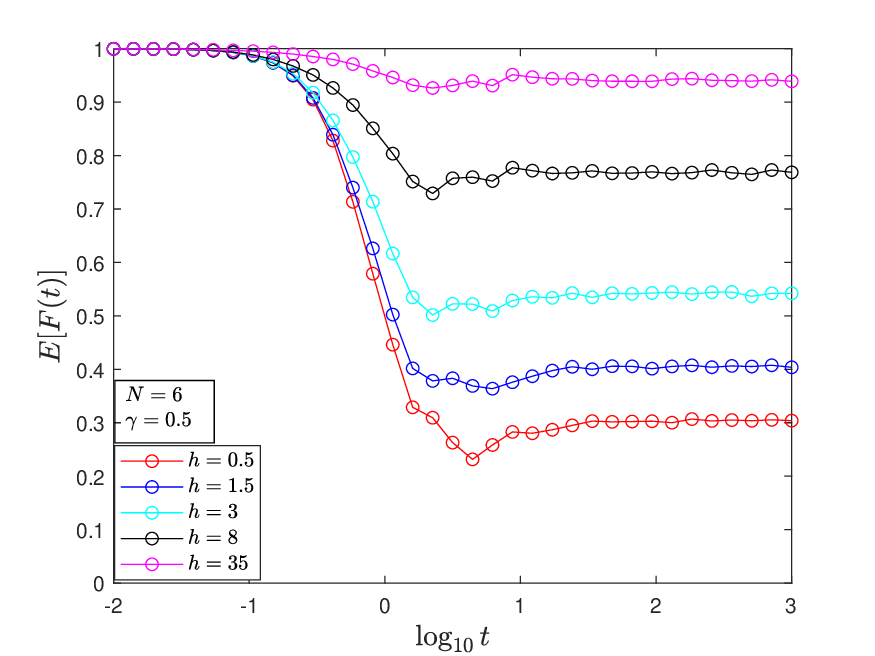}
\caption{The dynamical evolution of the averaged excited-state fidelity $\mathsf{E}[F(t)]$ for the anisotropic spin-1/2 chain with anisotropic all non-nearest-neighbor couplings at system size $N = 6$ and anisotropy parameter $\gamma = 0.5$. The legend indicates the values of disorder strength $h$.}
\label{fig12}
\end{figure}

It shows that, at low disorder strength ($h = 0.5$), fidelity stabilizes over time at a low value, indicating that the initial-state information is nearly lost and the system enters the thermalized phase, consistent with the eigenstate thermalization hypothesis. In contrast, at high disorder strength ($h = 35$), fidelity remains close to 1, signifying that the initial-state information is well preserved and the system is in the localized phase. As disorder strength increases, the decay of fidelity slows down, reflecting the transition from the ergodic phase to the localized phase.

By analyzing the dynamical behavior of diagonal entropy (DE), local magnetization, and the time evolution of fidelity, we further verify the occurrence of the MBL phase transition in the one-dimensional anisotropic spin-1/2 chains and distinguish between the ergodic (thermal) phase and the localized phase. Furthermore, the results indicate that in the localized phase, the system can effectively preserve its initial information if the disorder strength is sufficiently large.

To further investigate the MBL properties of the one-dimensional anisotropic spin-1/2 chain, we introduce Dzyaloshinskii--Moriya (DM) interaction \cite{dzyaloshinsky1958thermodynamic,
moriya1960anisotropic} into all three models.The Hamiltonians of the three models with the inclusion of DM interaction are given as follows:
\begin{equation}
\begin{split}
\hat{H} = \sum\limits_{i = 1}^{N - 1}  \hat{S}_i^z  \hat{S}_{i + 1}^z + \sum\limits_{i = 1}^N {h_i} \hat{S}_i^z + \sum\limits_{i = 1}^{N - 2} [(1 + \gamma)  \hat{S}_i^x  \hat{S}_{i + 2}^x + (1 - \gamma)  \hat{S}_i^y  \hat{S}_{i + 2}^y ] \\
+\sum\limits_{i = 1}^{N - 1}D\left(\hat{S}_{i}^{x} \hat{S}_{i+1}^{y}-\hat{S}_{i}^{y} \hat{S}_{i+1}^{x}\right),
\end{split}
\end{equation}

\begin{equation}
\hat{H} = \sum\limits_{i = 1}^{N - 1}  \hat{S}_i^z  \hat{S}_{i + 1}^z + \sum\limits_{i = 1}^N {h_i} \hat{S}_i^z +\gamma  \sum\limits_{i = 1}^{N }  \hat{S}_i^x +\sum\limits_{i = 1}^{N - 1}D\left(\hat{S}_{i}^{x} \hat{S}_{i+1}^{y}-\hat{S}_{i}^{y} \hat{S}_{i+1}^{x}\right),
\end{equation}

\begin{equation}
\begin{split}
\hat{H} =\sum\limits_{i = 1}^{N - 1}  \hat{S}_i^z  \hat{S}_{i + 1}^z + \sum\limits_{i = 1}^N {h_i} \hat{S}_i^z +\sum_{i=1}^{N-2} \sum_{j>i+1}^{N} \frac{1}{(j-i)^{\gamma}}\left[(1-\gamma) \hat{S}_{i}^{y} \hat{S}_{j}^{y}+(1+\gamma)  \hat{S}_{i}^{x} \hat{S}_{j}^{x}\right]\\
+\sum\limits_{i = 1}^{N - 1}D\left(\hat{S}_{i}^{x} \hat{S}_{i+1}^{y}-\hat{S}_{i}^{y} \hat{S}_{i+1}^{x}\right),
\end{split}
\end{equation}

We investigate the effect of DM interaction on the MBL phase transition in three one-dimensional anisotropic spin-1/2 chains, with system sizes $N = 6$ and $N = 8$. In Fig. 13, Fig. 14, and Fig. 15, we plot the averaged excited-state fidelity $\mathsf{E}[F]$ as a function of disorder strength $h$ for DM interaction strengths $D = 0$, 1, and 5, with the anisotropy parameter fixed at $\gamma = 0.5$.

\begin{figure}
    \centering
    \begin{subfigure}[b]{0.50\textwidth}
      \includegraphics[width=\textwidth]{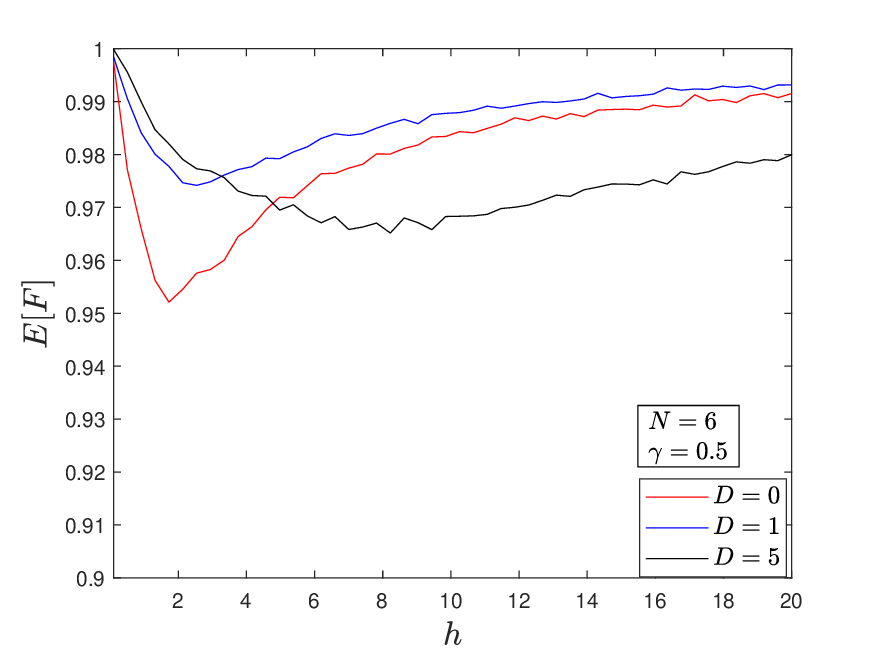}
      \caption{}
      \label{fig:oaspl_a}
    \end{subfigure}%
    ~
    \begin{subfigure}[b]{0.50\textwidth}
      \includegraphics[width=\textwidth]{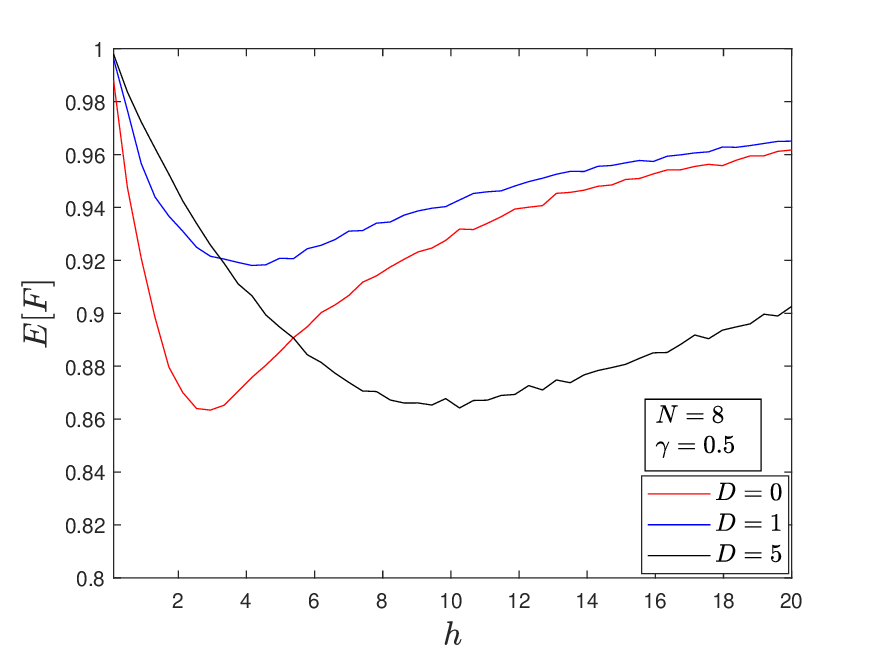}
      \caption{}
      \label{fig:oaspl_b}
    \end{subfigure}

  \caption{The averaged excited-state fidelity $\mathsf{E}[F]$ as a function of disorder strength $h$ for the anisotropic spin-1/2 chain with anisotropic next-nearest-neighbor coupling and DM interaction. (a) $\gamma = 0.5$, $N = 6$. (b) $\gamma = 0.5$, $N = 8$. The legend indicates the values of DM interaction strength $D$.}
    \label{fig13}
    \end{figure}

\begin{figure}
    \centering
    \begin{subfigure}[b]{0.50\textwidth}
      \includegraphics[width=\textwidth]{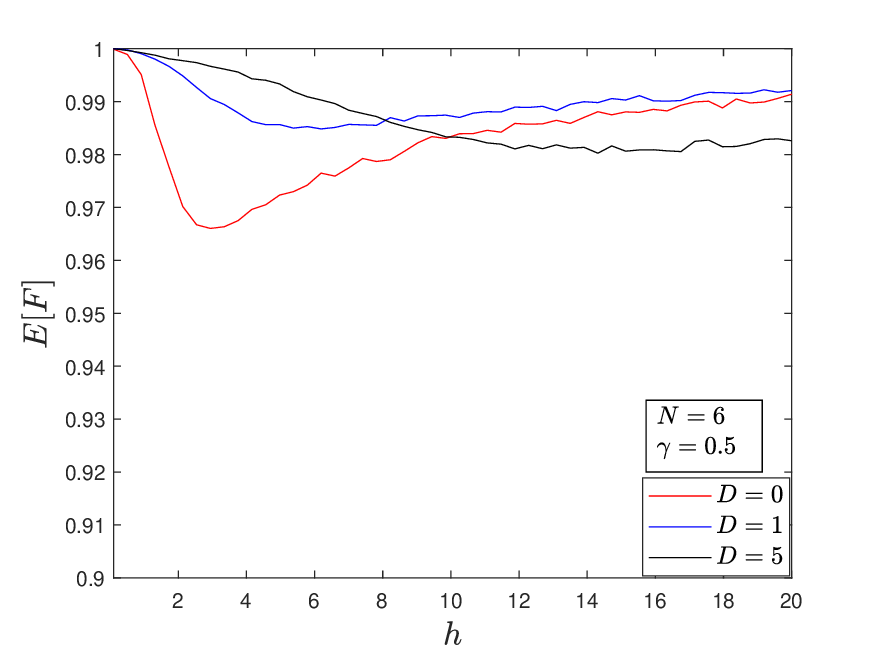}
      \caption{}
      \label{fig:oaspl_a}
    \end{subfigure}%
    ~
    \begin{subfigure}[b]{0.50\textwidth}
      \includegraphics[width=\textwidth]{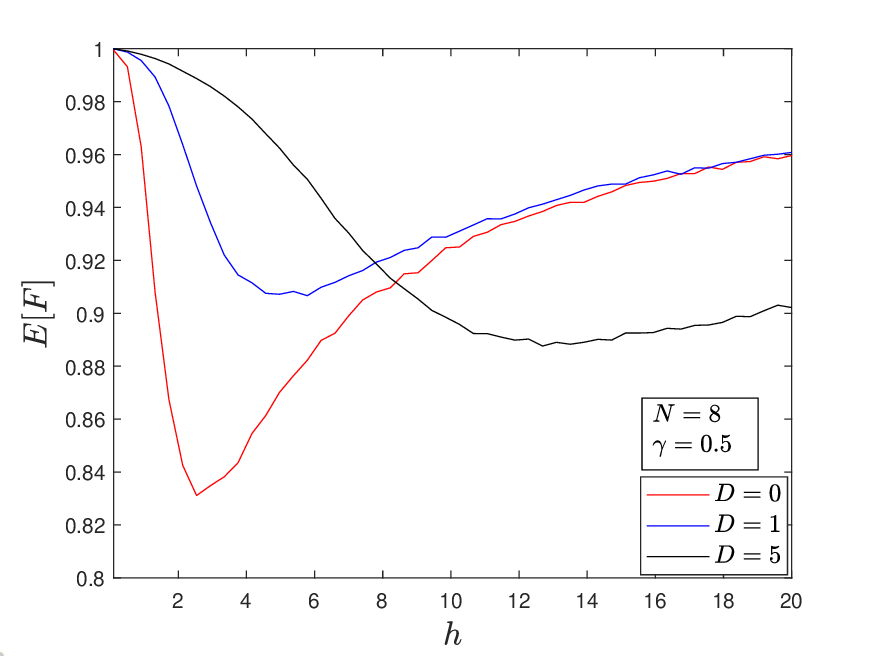}
      \caption{}
      \label{fig:oaspl_b}   
    \end{subfigure}
   
  \caption{The averaged excited-state fidelity $\mathsf{E}[F]$ as a function of disorder strength $h$ for the anisotropic spin-1/2 chain with an anisotropic external field and DM interaction. (a) $\gamma = 0.5$, $N = 6$. (b) $\gamma = 0.5$, $N = 8$. The legend indicates the values of DM interaction strength $D$.}
    \label{fig14}
    \end{figure}

\begin{figure}
    \centering
    \begin{subfigure}[b]{0.50\textwidth}
      \includegraphics[width=\textwidth]{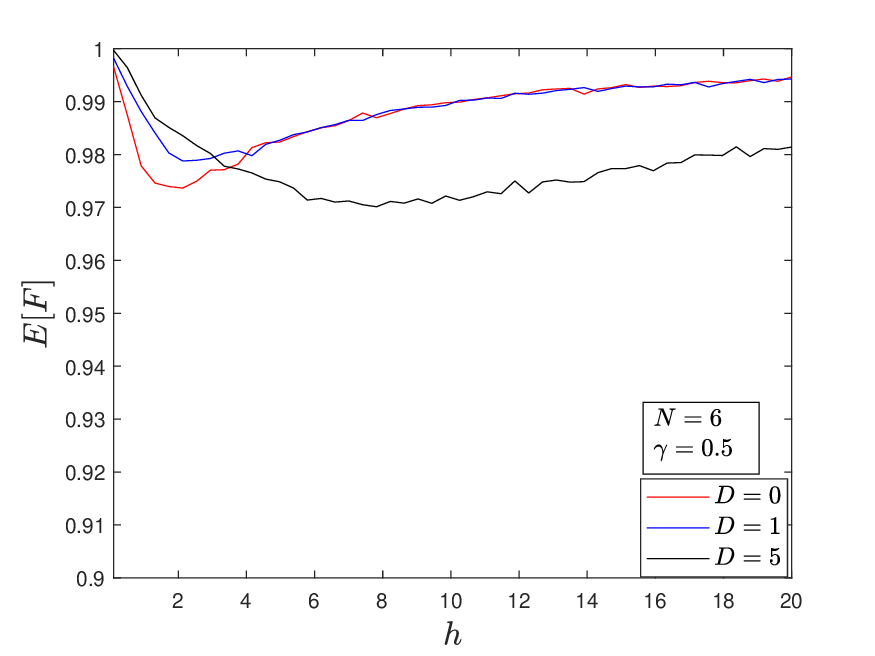}
      \caption{}
      \label{fig:oaspl_a}
    \end{subfigure}%
    ~
    \begin{subfigure}[b]{0.50\textwidth}
      \includegraphics[width=\textwidth]{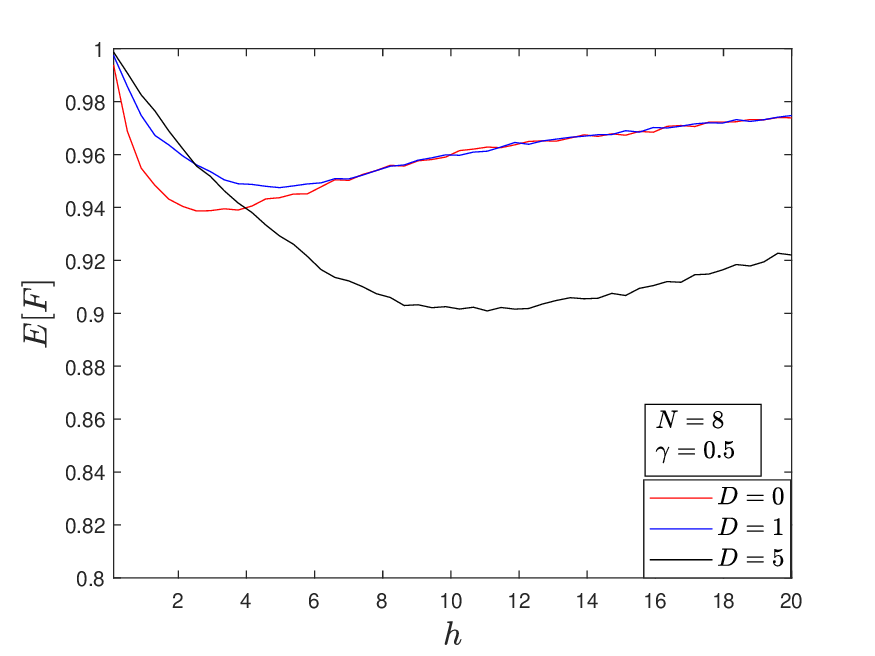}
      \caption{}
      \label{fig:oaspl_b}
    \end{subfigure}
   
  \caption{The averaged excited-state fidelity $\mathsf{E}[F]$ as a function of disorder strength $h$ for the anisotropic spin-1/2 chain with anisotropic all non-nearest-neighbor couplings and DM interaction. (a) $\gamma = 0.5$, $N = 6$. (b) $\gamma = 0.5$, $N = 8$.}
        \label{fig15}
    \end{figure}

The results indicate that the three one-dimensional anisotropic spin-1/2 chains with DM interaction can undergo the MBL phase transition. As shown in the figures, the DM interaction strength $D$ has an effect on the critical point of the MBL transition. Compared with the case without DM interaction ($D = 0$), the critical point becomes larger, and as $D$ increases, the critical disorder strength $h_c$ for the MBL transition also increases.

\begin{figure}
\centering
\includegraphics[width=0.70\textwidth]{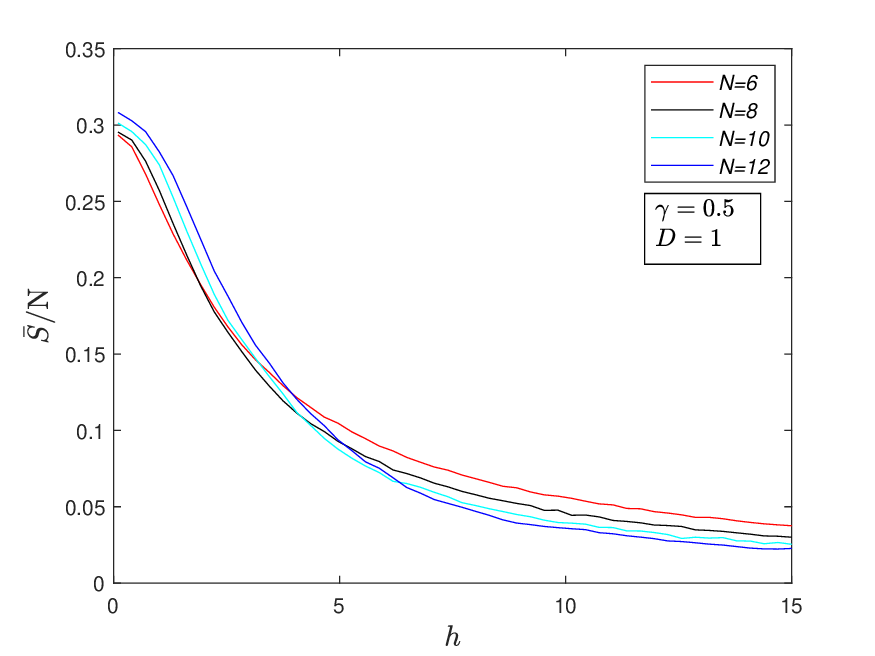}
\caption{The averaged bipartite entanglement entropy over the system size $\bar{S}/N$ for the anisotropic spin-1/2 chain with anisotropic next-nearest-neighbor couplings and DM interaction at fixed parameters $\gamma = 0.5$ and $D = 1$, as a function of the disorder strength $h$. The system sizes are indicated in the legend.}
\label{fig16}
\end{figure}

\begin{figure}
\centering
\includegraphics[width=0.70\textwidth]{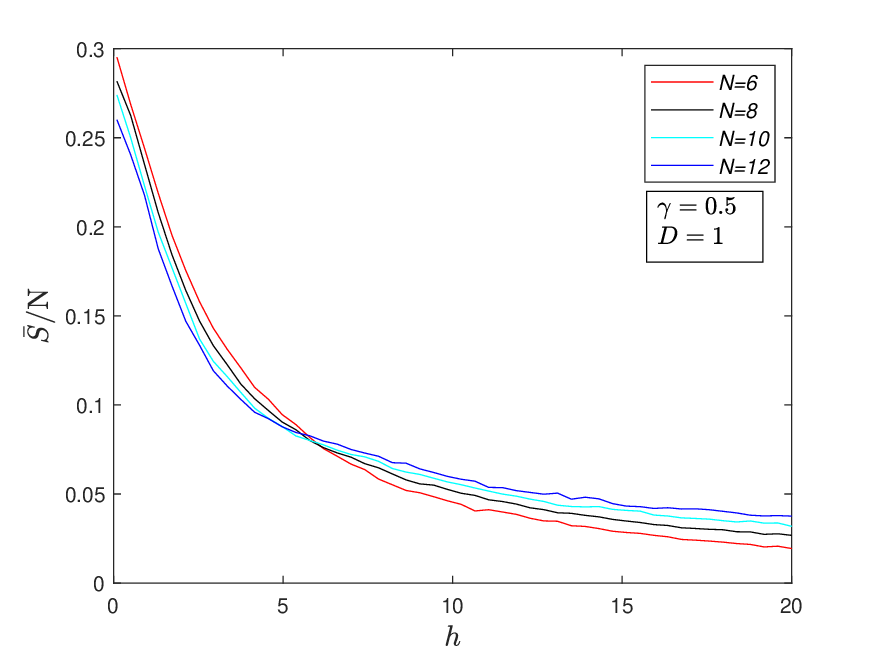}
\caption{The averaged bipartite entanglement entropy over the system size $\bar{S}/N$ for the anisotropic spin-1/2 chain with anisotropic external fields and DM interaction at fixed parameters $\gamma = 0.5$ and $D = 1$, as a function of the disorder strength $h$. The system sizes are indicated in the legend.}
\label{fig17}
\end{figure}

\begin{figure}
\centering
\includegraphics[width=0.70\textwidth]{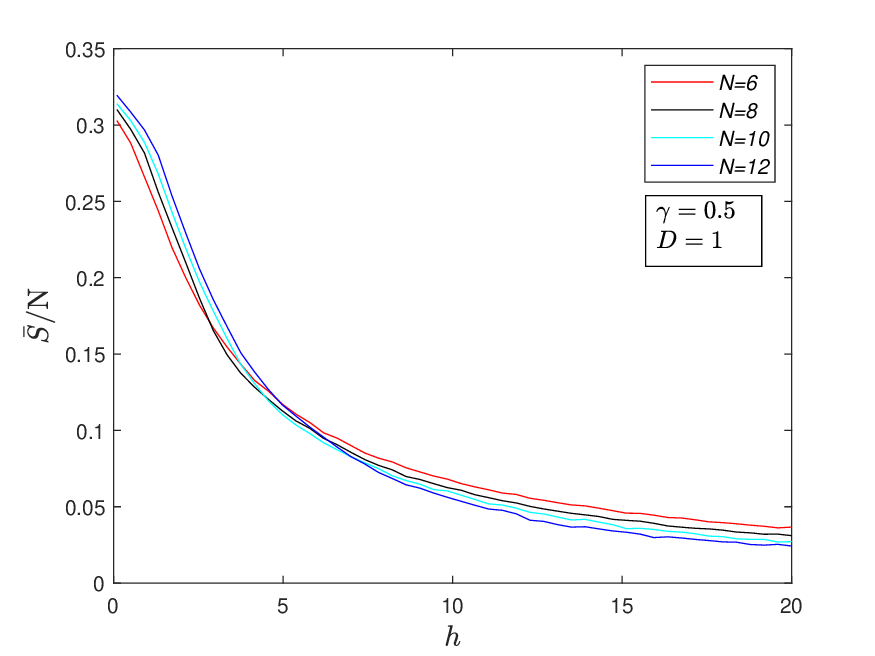}
\caption{The averaged bipartite entanglement entropy over the system size $\bar{S}/N$ for the anisotropic spin-1/2 chain with anisotropic all non-nearest-neighbor couplings and DM interaction at fixed parameters $\gamma = 0.5$ and $D = 1$, as a function of the disorder strength $h$. The system sizes are indicated in the legend.}
\label{fig18}
\end{figure}
By averaging over the selected excited eigenstates and all disorder realizations, we obtain the mean bipartite entanglement entropy $\bar{S}$. Figures 16, 17, and 18 illustrate the behavior of the normalized bipartite entanglement entropy $\bar{S}/N$ as a function of disorder strength $h$ for three anisotropic spin-1/2 models with DM interaction: (i) models with anisotropic next-nearest-neighbor couplings, (ii) models with anisotropic external fields, and (iii) models with all anisotropic non-nearest-neighbor couplings, respectively. The parameter is fixed at  $\gamma = 0.5$ and $D = 1$.

In the weak disorder regime, the value of $\bar{S}/N$ increases with the system size $N$ , which is consistent with the volume law and indicates strong quantum entanglement in the thermal phase. As $h$ increases, the entropy curves for different system sizes intersect in pairs. When $h$ becomes sufficiently large, $\bar{S}/N$ drops sharply toward zero, suggesting the system enters a localized phase and the entanglement in energy eigenstates is significantly suppressed.

\begin{figure}[!htbp]
    \centering
    \begin{subfigure}[b]{0.48\textwidth}
      \includegraphics[width=\textwidth]{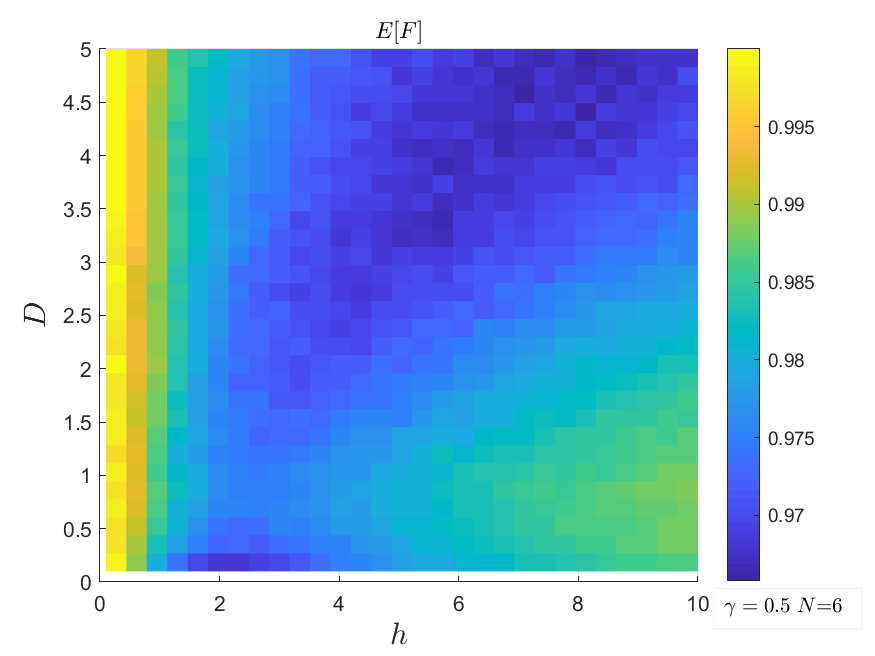}
      \caption{The averaged excited-state fidelity $\mathsf{E}[F]$ as a function of disorder strength $h$ and DM interaction strength $D$ at fixed parameters $\gamma = 0.5$ and $N = 6$.}
      \label{fig:oaspl_a}
    \end{subfigure}%
    ~
    \begin{subfigure}[b]{0.48\textwidth}
      \includegraphics[width=\textwidth]{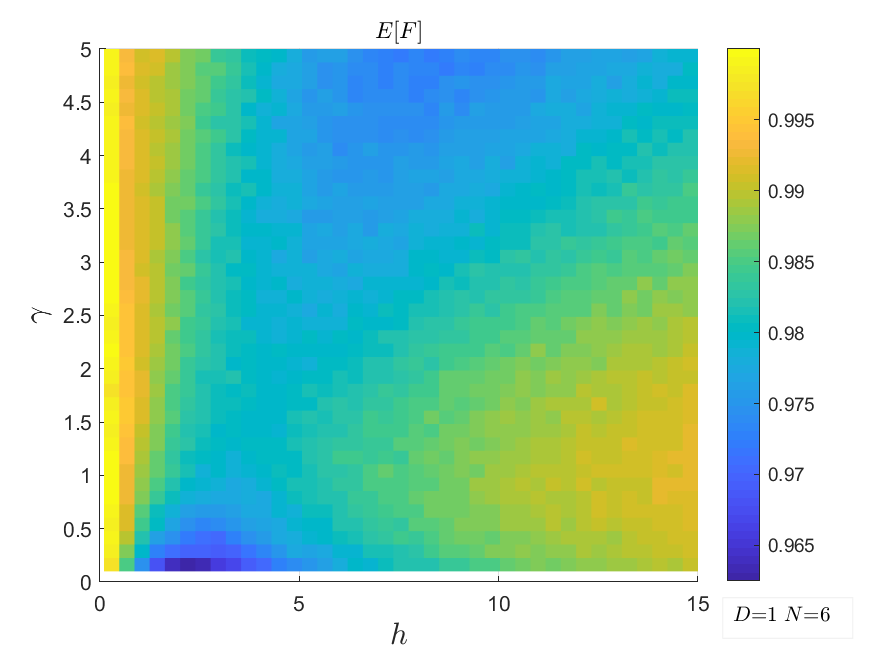}
      \caption{The averaged excited-state fidelity $\mathsf{E}[F]$ as a function of disorder strength $h$ and anisotropy parameter $\gamma$ at fixed parameters $D = 1$ and $N = 6$.}
      \label{fig:oaspl_b}
    \end{subfigure}
    \\
    \begin{subfigure}[b]{0.48\textwidth}
      \includegraphics[width=\textwidth]{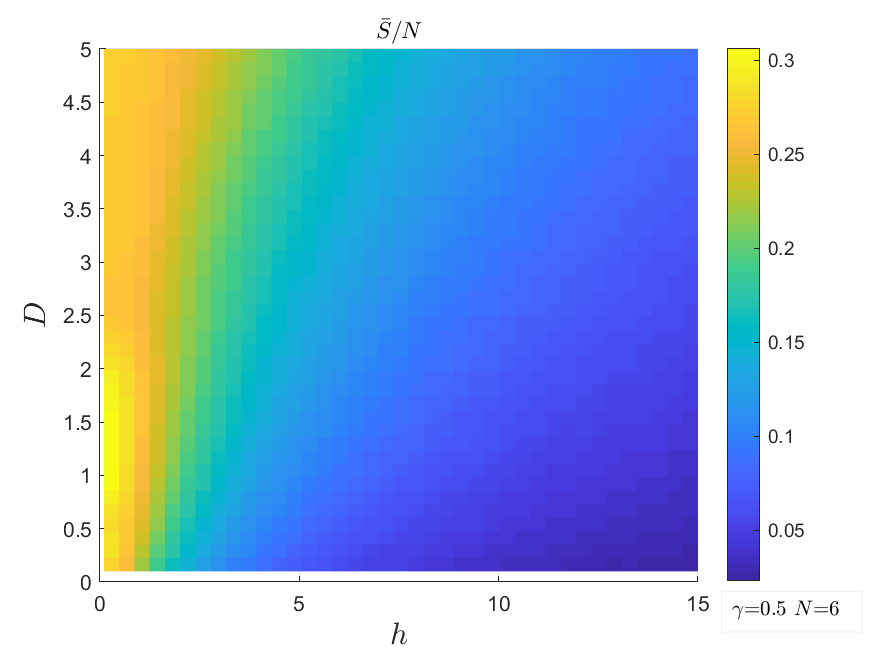}
      \caption{The averaged bipartite entanglement entropy per site $\bar{S}/N$ as a function of disorder strength $h$ and DM interaction strength $D$ at fixed parameters $\gamma = 0.5$ and $N = 6$.}
      \label{fig:oaspl_c}
    \end{subfigure}%
    ~
    \begin{subfigure}[b]{0.48\textwidth}
      \includegraphics[width=\textwidth]{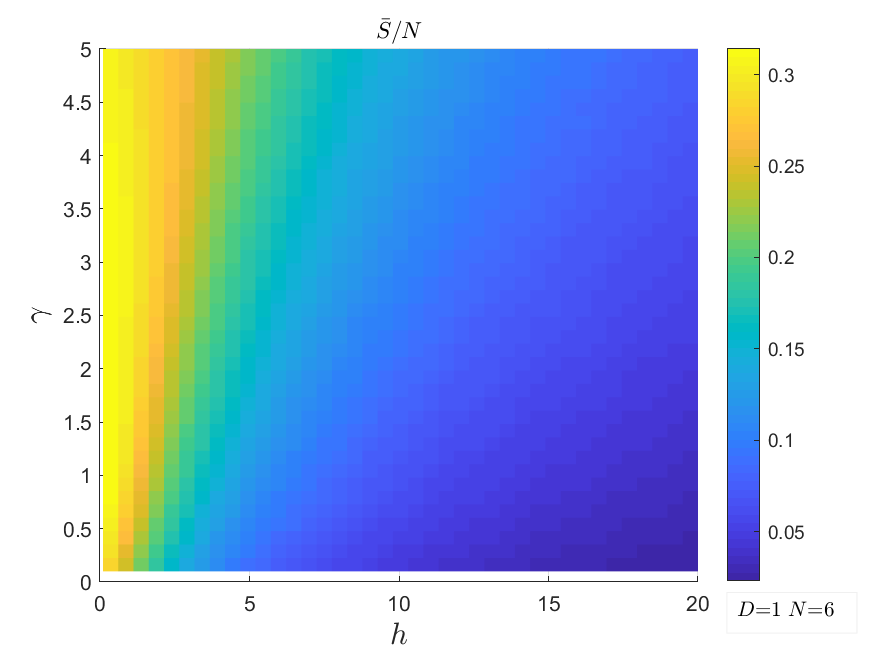}
      \caption{The averaged bipartite entanglement entropy per site $\bar{S}/N$ as a function of disorder strength $h$ and anisotropy parameter $\gamma$ at fixed parameters $D = 1$ and $N = 6$.}
      \label{fig:oaspl_d}
    \end{subfigure}
   \caption{Parameter-dependent behaviors of the average excited-state fidelity and bipartite entanglement entropy for the anisotropic spin-1/2 chain with anisotropic next-nearest-neighbor couplings and DM interaction. All results are computed at system size $N = 6$ under varying $h$, $D$, and $\gamma$.}
    \label{fig19}
    \end{figure}

\begin{figure}[!htbp]
    \centering
    \begin{subfigure}[b]{0.48\textwidth}
      \includegraphics[width=\textwidth]{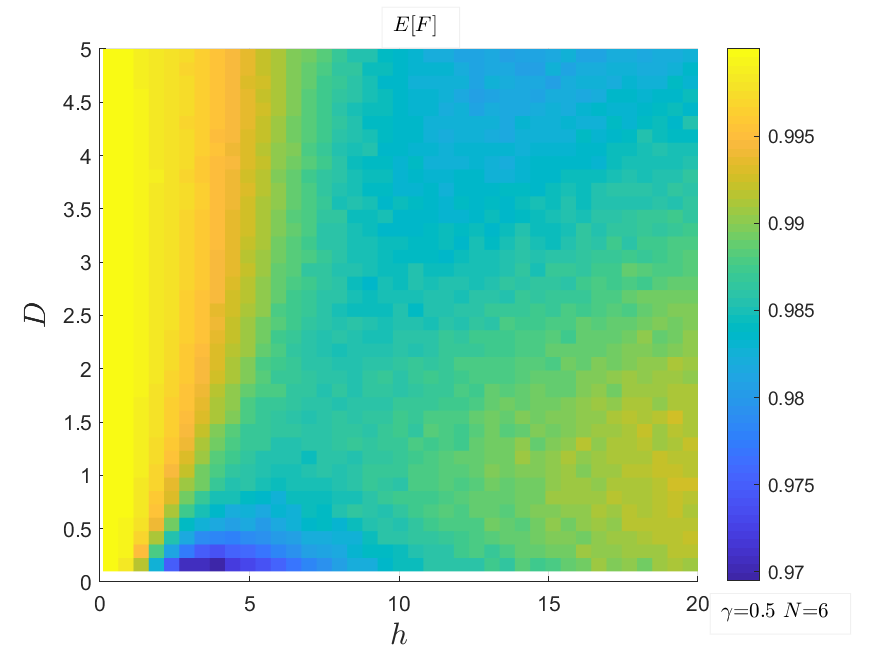}
      \caption{The averaged excited-state fidelity $\mathsf{E}[F]$ as a function of disorder strength $h$ and DM interaction strength $D$ at fixed parameters $\gamma = 0.5$ and $N = 6$.}
      \label{fig:oaspl_a}
    \end{subfigure}%
    ~
    \begin{subfigure}[b]{0.48\textwidth}
      \includegraphics[width=\textwidth]{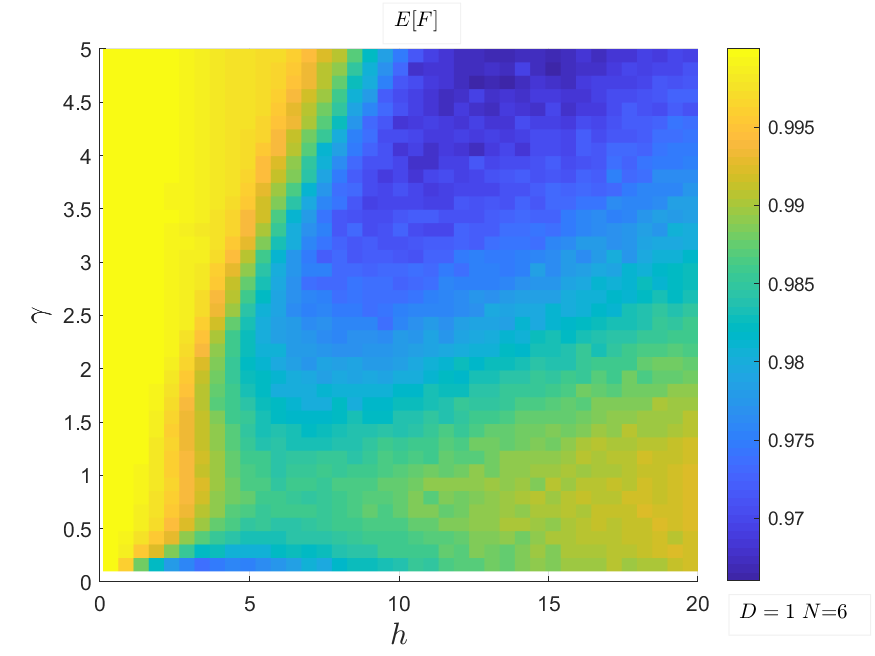}
      \caption{The averaged excited-state fidelity $\mathsf{E}[F]$ as a function of disorder strength $h$ and anisotropy parameter $\gamma$ at fixed parameters $D = 1$ and $N = 6$.}
      \label{fig:oaspl_b}
    \end{subfigure}
    \\
    \begin{subfigure}[b]{0.48\textwidth}
      \includegraphics[width=\textwidth]{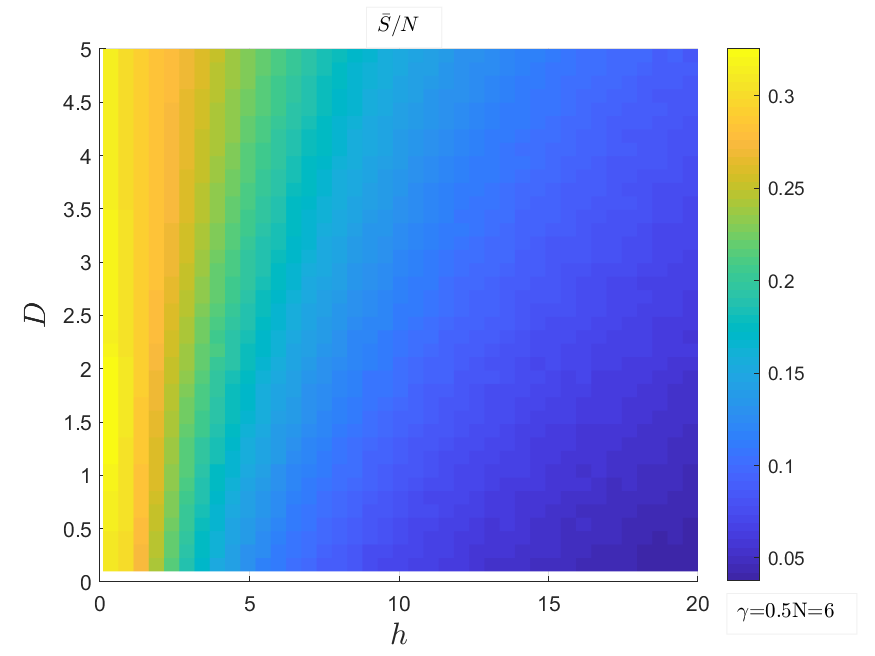}
      \caption{The averaged bipartite entanglement entropy per site $\bar{S}/N$ as a function of disorder strength $h$ and DM interaction strength $D$ at fixed parameters $\gamma = 0.5$ and $N = 6$.}
      \label{fig:oaspl_c}
    \end{subfigure}%
    ~
    \begin{subfigure}[b]{0.48\textwidth}
      \includegraphics[width=\textwidth]{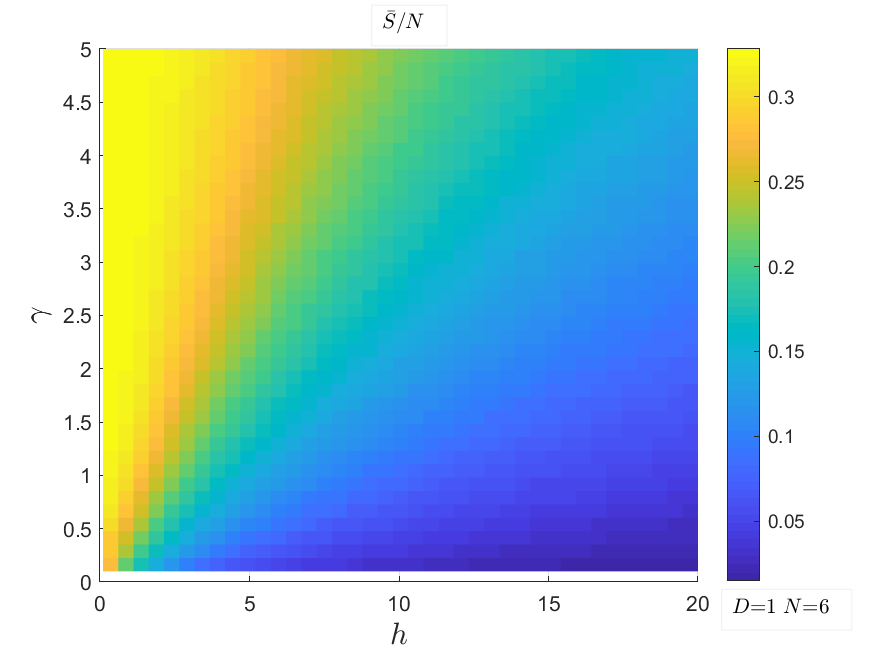}
      \caption{The averaged bipartite entanglement entropy per site $\bar{S}/N$ as a function of disorder strength $h$ and anisotropy parameter $\gamma$ at fixed parameters $D = 1$ and $N = 6$.}
      \label{fig:oaspl_d}
    \end{subfigure}
   \caption{Parameter-dependent behaviors of the averaged excited-state fidelity and bipartite entanglement entropy for the anisotropic spin-1/2 chain with anisotropic external fields and  DM interaction. All results are computed at system size $N = 6$ under varying $h$, $D$, and $\gamma$.}
    \label{fig20}
    \end{figure}

\begin{figure}[!htbp]
    \centering
    \begin{subfigure}[b]{0.48\textwidth}
      \includegraphics[width=\textwidth]{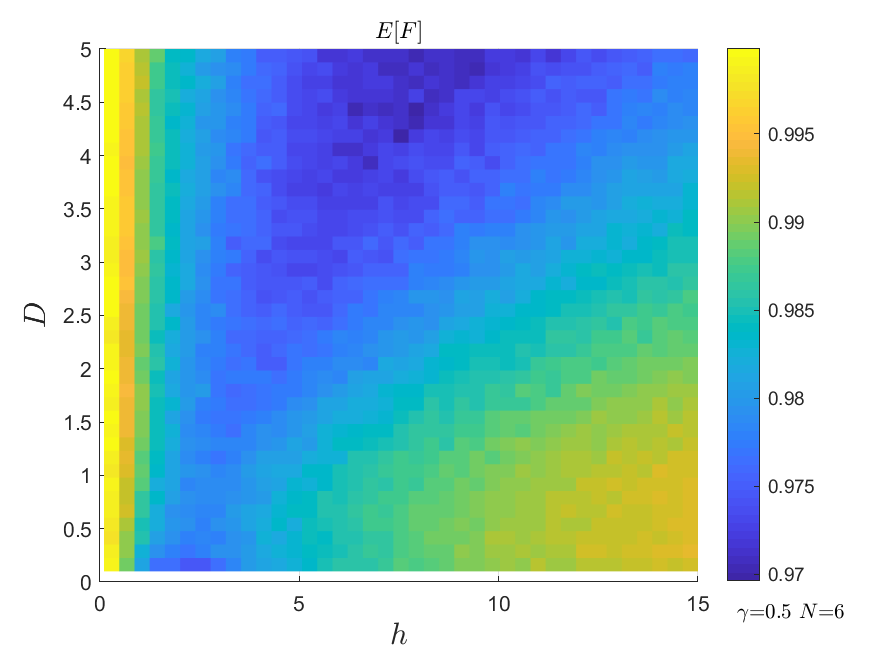}
      \caption{The averaged excited-state fidelity $\mathsf{E}[F]$ as a function of disorder strength $h$ and DM interaction strength $D$ at fixed parameters $\gamma = 0.5$ and $N = 6$.}
      \label{fig:oaspl_a}
    \end{subfigure}%
    ~
    \begin{subfigure}[b]{0.48\textwidth}
      \includegraphics[width=\textwidth]{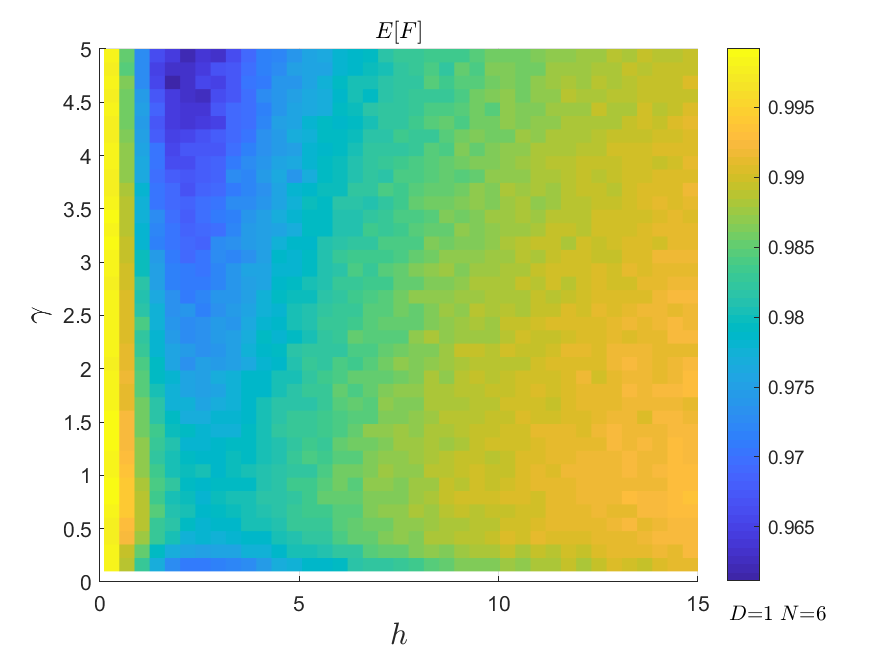}
      \caption{The averaged excited-state fidelity $\mathsf{E}[F]$ as a function of disorder strength $h$ and anisotropy parameter $\gamma$ at fixed parameters $D = 1$ and $N = 6$.}
      \label{fig:oaspl_b}
    \end{subfigure}
    \\
    \begin{subfigure}[b]{0.48\textwidth}
      \includegraphics[width=\textwidth]{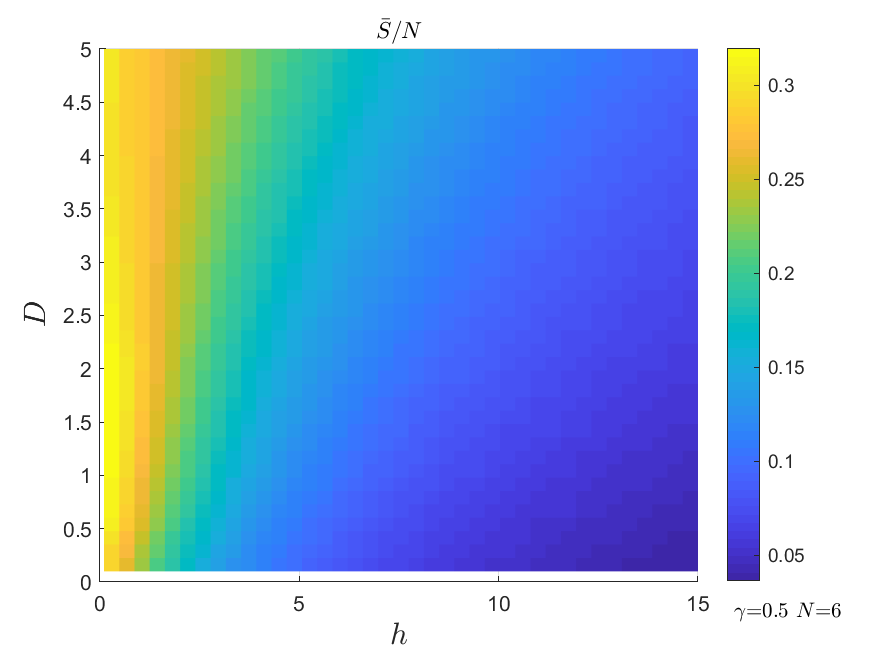}
      \caption{The averaged bipartite entanglement entropy per site $\bar{S}/N$ as a function of disorder strength $h$ and DM interaction strength $D$ at fixed parameters $\gamma = 0.5$ and $N = 6$.}
      \label{fig:oaspl_c}
    \end{subfigure}%
    ~
    \begin{subfigure}[b]{0.48\textwidth}
      \includegraphics[width=\textwidth]{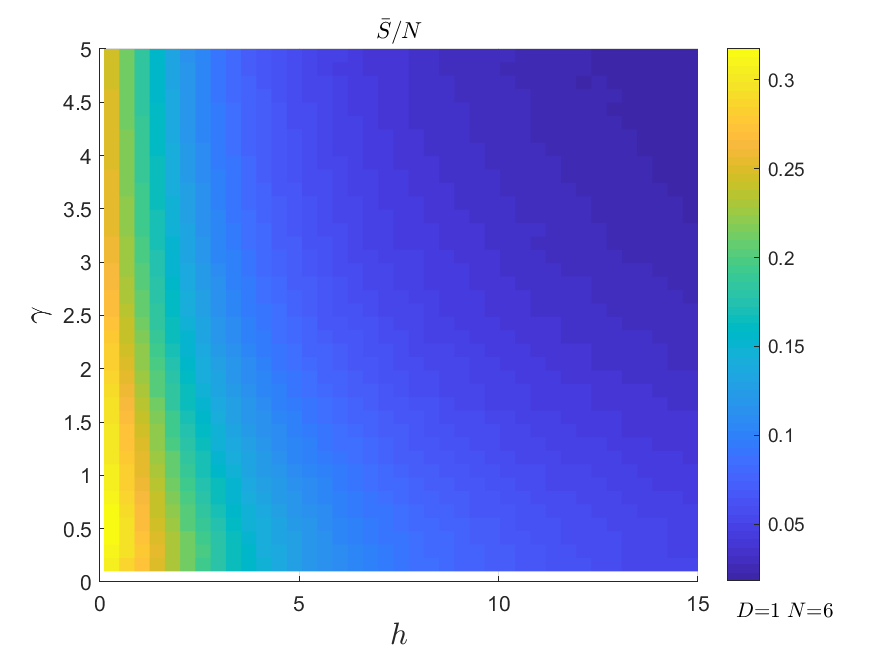}
      \caption{The averaged bipartite entanglement entropy per site $\bar{S}/N$ as a function of disorder strength $h$ and anisotropy parameter $\gamma$ at fixed parameters $D = 1$ and $N = 6$.}
      \label{fig:oaspl_d}
    \end{subfigure}
   \caption{Parameter-dependent behaviors of the averaged excited-state fidelity and bipartite entanglement entropy for the anisotropic spin-1/2 chain with anisotropic all non-nearest-neighbor couplings and DM interaction. All results are computed at system size $N = 6$ under varying $h$, $D$, and $\gamma$.}
    \label{fig21}
    \end{figure}

Similarly, based on the intersection points of the critical system size N curves, we can obtain the phase transition critical point:

\begin{itemize}
    \item In the model with anisotropic next-nearest-neighbor couplings and DM interaction,  the critical point is approximately $h_c \approx 1.9$ for $N=6$, $h_c \approx 4.1$ for $N=8$, and $ h_c \approx 6.3 $ for $N=10$, showing an increasing trend of the transition point with system size.
    
    \item In the model with anisotropic external fields and DM interaction, the intersection points are found at $h_c \approx 6.1$, $5.8$, and $5$, suggesting a slight decrease in the critical point as the system size increases.
    
    \item In the model with all anisotropic non-nearest-neighbor couplings and DM interaction, the critical points are located at $h_c \approx 2.5$, $4.4$, and $7$, showing a clear increasing trend with system size.
\end{itemize}

These critical values are also highly consistent with those obtained from the analysis of excited-state fidelity, confirming that all three anisotropic spin-1/2 chain models with DM interaction undergo many-body localization transitions. Furthermore, the results provide reliable estimates for the MBL critical points in each system.
   
Since research has shown that both the anisotropic parameter $\gamma$ and the DM interaction strength $D$ affect the critical point of the MBL phase transition in the three models with DM interaction, we therefore plotted the averaged excited-state fidelity $\mathsf{E}[F]$ and the ratio of the averaged bipartite entanglement entropy to the system size $\bar{S}/N$ as functions of $h$ and $D$ for system size $N = 6$, as shown in panels (a) and (c) of Figures 19, 20, and 21; as well as their variations with respect to $h$ and $\gamma$, as shown in panels (b) and (d) of the same figures.

As shown in subfigures (b) and (d) of Figures 19 and 20, when the DM interaction strength \( D \) is fixed, the critical disorder strength \( h_c \) for the MBL transition increases with the anisotropy parameter \( \gamma \) in the anisotropic spin-1/2 chains with anisotropic next-nearest-neighbor couplings and anisotropic external fields under the presence of DM interaction.

In contrast, a different trend is observed in subfigures (b) and (d) of Figure 21: when \( D \) is fixed, the critical disorder strength \( h_c \) gradually decreases with increasing \( \gamma \) in the anisotropic spin-1/2 chain with all non-nearest-neighbor couplings. Moreover, the decrease becomes less pronounced as \( \gamma \) increases.

Furthermore, subfigures (a) and (c) of Figures 19, 20, and 21 reveal that when the anisotropy parameter \( \gamma \) is fixed, the critical disorder strength \( h_c \) for the MBL transition shows a rising trend with increasing DM interaction strength \( D \) across all three models.

\section{Summary }

In this paper, we investigate the phenomenon of many-body localization (MBL) in one-dimensional anisotropic spin-1/2 chains using the exact matrix diagonalization method. Three different systems are considered: the anisotropic spin-1/2 chain with next-nearest-neighbor coupling, with an anisotropic external field, and with all non-nearest-neighbor couplings. By computing the averaged excited-state fidelity, we analyze the eigenstate properties of these systems and estimate the critical points of the MBL phase transition based on the characteristic changes in the fidelity curves. The results show that both the anisotropy parameter and system size affect the critical point of the transition.Furthermore, the bipartite entanglement entropy was calculated for different system sizes, and the resulting critical points were found to be largely consistent with those obtained from the fidelity analysis, confirming the occurrence of the MBL phase transition.
Subsequently, we study the dynamical properties of the systems through the time evolution of diagonal entropy (DE), local magnetization, and fidelity. The results further confirm the occurrence of the MBL phase transition in the one-dimensional anisotropic spin-1/2 chains and clearly distinguish the thermal (ergodic) phase from the many-body localized phase. It is also shown that in the localized phase, if the disorder strength is sufficiently large, the system can effectively preserve its initial information.
To further explore the MBL properties in these systems, we introduce DM interaction into the three models. The results indicate that DM interaction, to some extent, suppresses the occurrence of the MBL phase transition. We hope that this study can provide meaningful insights and references for further investigations into the phenomenon of many-body localization.

\section*{Acknowledgments}
This work was supported by the Plan for Scientific and Technological Development of Jilin Province,
(No. 20230101018JC), by the NSF of China (Grant No.
62175233), by the NSF of China (Grant No. 62375259), and
by the Plan for Scientific and Technological Development of
Jilin Province (No. 20220101111JC).

\bibliographystyle{unsrt}
\bibliography{xxx1}  

\end{document}